\documentclass[11pt,a4paper,epsfig]{article}

\usepackage{amsfonts}
\usepackage{graphicx}
\usepackage[colorlinks=true,linkcolor=blue]{hyperref}

\title{\textbf{On a family of (1+1)-dimensional scalar field theory models: kinks, stability, one-loop mass shifts}}

\author{A. Alonso-Izquierdo$^{(a)}$, and J. Mateos Guilarte$^{(b)}$
\\ {\normalsize {\it $^{(a)}$ Departamento de Matematica
Aplicada and IUFFyM}, {\it Universidad de Salamanca, SPAIN}} \\ {\normalsize {\it $^{(a)}$ Departamento de Fisica
Fundamental and IUFFyM}, {\it Universidad de Salamanca, SPAIN}}}

\begin{document}

\maketitle

\begin{abstract}
In this paper we construct a one-parametric family of (1+1)-dimensional one-component scalar field theory models supporting kinks. Inspired by the sine-Gordon and $\phi^4$ models, we look at all possible extensions such that the kink second-order fluctuation operators are Schr\"odinger differential operators with P\"oschl-Teller potential wells. In this situation, the associated spectral problem is solvable and therefore we shall succeed in analyzing the kink stability completely and in computing the one-loop quantum correction to the kink mass exactly. When the parameter is a natural number, the family becomes the hierarchy for which the potential wells are reflectionless, the two first levels of the hierarchy being the sine-Gordon and $\phi^4$ models.
\end{abstract}

\section{Introduction}

From the very beginning of the surge of solitons in Quantum Field Theory in the mid seventies it was noticed that the second-order kink fluctuation operators in the sine-Gordon and $\phi^4$ models were respectively the two first members in the hierarchy of Schr\"odinger operators with \lq\lq transparent" P\"oschl-Teller wells:
\[
K^{(1)} = -\frac{d^2}{dx^2}+1 - 2\,{\rm sech}^2 x  \hspace{1cm},\hspace{1cm} K^{(2)} = -\frac{d^2}{dx^2}+4 - 6\,{\rm sech}^2 x \qquad .
\]
In 1975, Christ and Lee in Reference \cite{Christ1975} addressed the following natural problem: what should the potential energy densities in the parent field theoretical models be such that the kink second-order fluctuation operators are the other members in the \lq\lq transparent" PT hierarchy: $K^{(N)} = -\frac{d^2}{dx^2}+N^2 - N(N+1)\,{\rm sech}^2 x$, $N\in{\mathbb N}$? These authors found the general answer and described some analytical properties of the $\phi_K^{(N)}$-kinks and the $U^{(N)}(\phi)$-potentials arising in this kind of model. They also noticed that the parent potentials had branch points at the vacua for $N\geq 3$. In 1987, Trullinger and Flesch \cite{Trullinger1987} succeeded in constructing the $N\in \mathbb{N}$ hierarchy by identifying the parent potentials in terms of some specific special functions. These functions are well defined when the field takes values in the interval in the field space between the two vacua. Beyond those points they tried to build the parent potentials by sewing the special functions just mentioned to some well-behaved functions with suitable properties at higher than the vacuum values of the field. The connection between the two zones, however, was at most twice differentiable at the branching points. Nevertheless, they showed that keeping only lowest-order (Gaussian) field fluctuations around the static vacua and kink solutions,  the parent potentials are well behaved. This left open the question of computing the one-loop mass quantum corrections to the kinks associated with these parent potentials. In 1989 Boya and Casahorran applied the procedure of Cahill, Comtet and Glauber \cite{Cahill1976} to the models in this hierarchy. Thus, they computed the one-loop kink mass quantum correction in terms of the bound state eigenvalues of the $K^{(N)}$ operator, see \cite{Boya1989}. We stress that this formula is valid only if the potential well is reflectionless and suitably integrable.

In a almost parallel development, the problem of the computation of the one-loop mass shifts for the sine-Gordon soliton and $\phi^4$ kink was addressed. The first to tackle this were Dashen, Hasslacher and Neveu in 1974 in the framework of the $\hbar$-expansion of these (1+1)-dimensional field theories \cite{Dashen1974,Dashen1975,Dashen1975b}. After a zero-point renormalization, which is performed by means of a mode-by-mode subtraction of the eigenvalues of the second-order fluctuation operators valued respectively on the kink and vacuum solutions and a subsequent mass renormalization, the authors derived the so-called first DHN formula. We stress that the DHN formula is again only valid when the kink second-order fluctuation operator is a reflectionless differential operator. In this case, both the kink and vacuum Hessians have a half-bound state whose contributions to the quantum correction annihilate each other in the mode-number cut-off regularization. In reference \cite{Alonso2004}, the authors construct a generalized DHN formula which works smoothly when applied to quite general Sch$\ddot{\rm o}$dinger operators. At the end of the past century a revival in the interest in this problem took place mainly due to r$\hat{\rm o}$le played by quantum solitons in supersymmetric theories. Many subtleties of the regularization/renormalization procedures involved were clarified, starting with the papers \cite{Rebhan1997, Nastase1999, Goldhaber2004, Rebhan2004}. Other related works,  \cite{Shifman1999,Graham1999,Graham1999b}, addressed mixed issues in the problem by using different types of boundary conditions - PBC, Dirichlet, Robin-, regularization methods -energy cutoff, mode-number cutoff, high-derivatives-, and/or performed phase-shift analysis, in connection with possible modifications due to the quantum effects of the central charge of the SUSY algebra \cite{Bordag1995,Bordag2002}. In parallel, topological defect quantum fluctuations
have been studied within the general framework of quantum vacuum fluctuations in recent years and the application of
spectral heat kernel/zeta function regularization methods in this kind of problems has grown very effectively, see \cite{Mateos2009}.

In the present paper, however, we address a two-fold goal:
\begin{enumerate}
\item We extend the analysis to the case where the parameter $\sigma$ of the P\"oschl-Teller well is a positive real number, rather than a natural number,
\begin{equation}
K^{(\sigma)} = -\frac{d^2}{dx^2}+\sigma^2 - \sigma(\sigma+1)\,{\rm sech}^2 x
\label{PoschlTeller}
\end{equation}
descending to the Christ-Lee-Trullinger-Flesch-Boya-Casahorran hierarchy when $\sigma=N \in\mathbb{N}$. We shall find analytically the kink solutions, which satisfy intriguing recurrence relations, in the form of regularized incomplete Beta functions.

\item We compute the one-loop kink mass shifts by a variety of methods, depending on the difficulty of the problem. For $\sigma=N\in \mathbb{N}$ the Cahill-Comtet-Glauber (CCG) formula \cite{Cahill1976} is applicable but we shall also compute the quantum correction by the Dashen-Hasslacher-Neveu (DHN) mode-number cut-off regularized formula \cite{Dashen1974} and/or the exact heat kernel/zeta function regularization method \cite{Alonso2002}. If $\sigma$ is not an integer, only the generalized DHN procedure as explained in \cite{Alonso2004} can be applied. Alternatively, the asymptotic expansion of the heat kernel will allow us to obtain sufficiently good approximations to the one-loop kink mass quantum correction.
\end{enumerate}

The organization of the paper is as follows: in Section \S. 2 we shall construct the family of models and we shall also identify the different kinks, discussing interesting relationships between the kinks in different models of the family. We shall also study kink stability, which by construction reduces to the spectral analysis of the operator (\ref{PoschlTeller}). In Section \S. 3 we shall compute the one-loop quantum correction to the kink masses. We shall distinguish the cases $\sigma\in \mathbb{N}$ and $\sigma \notin \mathbb{N}$, or equivalently, we shall deal with the Hessian operator (\ref{PoschlTeller}) with reflection and reflectionless potential well separately. We shall approach this calculation from different points of view. In particular, we shall use the CCG (Cahill, Comtet and Glauber) formula, the DHN (Dashen, Hasslacher and Neveu) formula, the spectral zeta function regularization method and the asymptotic approach. The first three methods can only be applied to models that involve reflectionless Hessian operators, although the second one has been generalized in reference \cite{Alonso2004}. A comparison between the results obtained with the different methods will be offered, showing a high level of accuracy. In the last Section we shall summarize our conclusions.

\section{The family of models: kinks and their stability}
\subsection{Generalities}
The action governing the dynamics in our (1+1)-dimensional relativistic one-scalar field theoretical models is of the form {\footnote{The conventions are those set out in reference \cite{Alonso2011}.}:
\[
\tilde{S}[\psi]=\int \!\! \int\, dy^0dy^1 \, \left(\frac{1}{2}\frac{\partial\psi}{\partial y_\mu}\cdot \frac{\partial\psi}{\partial y^\mu}- \tilde{U}[\psi(y^\mu)] \right) \quad .
\]
Here, $\psi(y^\mu): \mathbb{R}^{1,1} \rightarrow \mathbb{R}$ is a real scalar field; i.e., a continuous map from the $(1+1)$-dimensional Minkowski space-time to the field of the real numbers.  $y^0=\tau $ and $y^1=y$ are local coordinates in ${\mathbb R}^{1,1}$, which is equipped with a metric tensor $g_{\mu\nu}={\rm diag}(1,-1), \mu,\nu=0,1$ such that $y_\mu y^\mu=g^{\mu\nu}y_\mu y_\nu$, $\frac{\partial}{\partial y_\mu}.\frac{\partial}{\partial y^\mu}=g^{\mu\nu}\frac{\partial}{\partial y_\mu}.\frac{\partial}{\partial y_\nu}$.

We shall work in a system of units where the speed of light is set to one, $c=1$, but we shall keep the Planck constant $\hbar$ explicit because we shall search for one-loop corrections, proportional to $\hbar$, to the classical kink masses.
In this system, the physical dimensions are:
\[
[\hbar]=[\tilde{S}]=M L \quad , \quad [y_\mu]=L \quad , \quad [\psi]=M^\frac{1}{2}L^\frac{1}{2} \quad , \quad [\tilde{U}]=ML^{-1} \quad .
\]
The models that we shall consider are distinguished by different choices of the part of the potential energy density which is independent on the field spatial derivatives: $\tilde{U}[\psi(y^\mu)]$. In all of them there will be two special parameters, $m_d$ and $\gamma_d$, to be determined in each case, carrying the physical dimensions: $[m_d]=L^{-1}$ and $[\gamma_d]=M^{-\frac{1}{2}}L^{-\frac{1}{2}}$. We define the non-dimensional coordinates, fields and potential in terms of these parameters:
\[
x_\mu=m_d y_\mu \, \, \, , \, \, \, x_0=t \, \, , \, \, x_1=x \quad , \quad \phi= \gamma_d \psi \qquad , \qquad U(\phi)=\frac{\gamma_d^2}{m_d^2} \tilde{U}(\psi) \qquad .
\]
The action $\tilde{S}[\psi]$ is also proportional to a non-dimensional action, i.e., $\tilde{S}[\psi]=\frac{1}{\gamma_d^2}S[\phi]$, where
\begin{equation}
S[\phi]=\int dx^2 \left[ \frac{1}{2} \partial_\mu \phi\, \partial^\mu \phi - U(\phi) \right] \qquad .\label{action}
\end{equation}
Here $\phi(x_\mu):\mathbb{R}^{1,1} \rightarrow \mathbb{R}$ is a non-dimensional real scalar.
Also, we define a non-dimensional potential energy functional by means of the relationship $E[\phi]$, where
\begin{equation}
E[\phi]=\frac{\gamma_d^2}{m_d}\tilde{E}[\psi]\quad , \quad E[\phi] = \int_{-\infty}^\infty \, dx \, \varepsilon(x) = \int_{-\infty}^\infty \, dx \, \left[ \frac{1}{2} \left( \frac{\partial \phi}{\partial x} \right)^2 + U(\phi) \right]
\quad .
\label{energy}
\end{equation}
From now on we shall work in this non-dimensional context. We shall assume that $U(\phi)$ is a non-negative twice continuously differentiable function of $\phi\in \mathbb{R}$, i.e., $U(\phi)\in C^2({\mathbb{R}})$ and $U(\phi)\geq 0$. We now want to construct kink supporting models where the kink stability is governed by P\"oschl-Teller spectral problems. We shall thus assume that the $U(\phi)$ has at least two minima -which can be fixed as the zeroes of $U(\phi)$- at the values $\phi=\pm 1$. This assumption does not imply a loss of generality because a translation of the origin and a dilatation in the field space allows us to set this prescription in an arbitrary model.

The configuration space ${\cal C}$ of the system is the set of the maps from the spatial line $\mathbb{R}$ to the internal (field) space $\mathbb{R}$ for a fixed time $t$, such that the energy functional $E[\phi]$ is finite, ${\cal C}=\{\phi(t,x)\in {\rm Maps}(\mathbb{R}^{1},\mathbb{R}):E[\phi]<\infty\}$. The search for kinks is tantamount to the search for finite energy static solutions $\phi(x)$ of the field equations, which in this case become the \lq\lq Newton" ODE:
\begin{equation}
\frac{d^2 \phi}{d x^2} = \frac{\partial U}{\partial \phi} \label{ode} \qquad .
\end{equation}
A Lorentz transformation shows the static solution $\phi(x)$ in its center of mass as the traveling wave: $\phi(t,x)=\phi(\frac{x-v t}{\sqrt{1-v^2}})$. Moreover, spatial translations and reflections, $x\rightarrow x+x_0$, $x\mapsto -x$, provide more solutions because of the invariance of (\ref{ode}) with respect to these transformations. Therefore, the complete family of solitary wave solutions of (\ref{ode}) is compactly written as: $\phi(\overline{x})$, $\overline{x}=(-1)^a \frac{x-x_0-vt}{\sqrt{1-v^2}}$ with $a=0,1$.

The second-order ODE (\ref{ode}) always admits the \lq\lq mechanical energy" $I_1=\frac{1}{2} \left( \frac{d\phi}{dx} \right)^2 - U(\phi)$ as a first integral. Because the \lq\lq mechanical action" is the field theoretical energy, the energy finiteness requirement identifies the kink solutions as the trajectories of $I_1=0$ mechanical energy:
\begin{equation}
\frac{d\phi}{dx} = \pm\sqrt{2 U(\phi)} \qquad .
\label{ode1}
\end{equation}
Therefore, the kink solutions interpolate between the $\phi_V=\mp 1$ vacuum at $x=-\infty$ and the $\phi_V=\pm 1$ vacuum at $x=+\infty$:
\[
\lim_{x\rightarrow -\infty} \phi_K(t,x)=\mp 1 \in {\cal M} \qquad , \qquad \lim_{x\rightarrow +\infty} \phi_K(t,x)=\pm 1 \in {\cal M} \qquad .
\]
From the signs in the equation (\ref{ode1}) one sees that the interpolation is monotonic: the kink solutions are either monotonically increasing (kinks) or decreasing (anti-kinks) functions of $x$ from $\mp 1$ at the left boundary of the line to $\pm 1$ on the opposite spatial limit.

The linear stability analysis of any static solution - vacua $\phi_V$ or kink $\phi_K(x)$- is codified by the spectral problems $K_0 f_{\omega_0}(x)= \omega_0^2 f_{\omega_0}(x)$ and $K f_\omega(x)= \omega^2 f_\omega(x)$ for the associated second-order fluctuation (Hessian) operators:
\begin{equation}
K_0  = -\frac{d^2}{dx^2} + \frac{\partial^2 U}{\partial \phi^2}[\phi_V] \quad , \quad K  = -\frac{d^2}{dx^2} + \frac{\partial^2 U}{\partial \phi^2}[\phi_K(x)] \quad .
\label{hessianoV}
\end{equation}
The stability of the solutions $\phi_V$ and $\phi_K(x)$ is guaranteed when the eigenvalues of these operators are non-negative. In particular, we are interested in the \lq\lq parent" potentials that admit kink solutions such that:
\begin{equation}
K^{(\sigma)}= -\frac{d^2}{dx^2} + \sigma^2- \sigma  (\sigma  +1) \, {\rm sech}^2 \overline{x}
\label{PoschtTellerPotential} \quad ,
\end{equation}
where $\sigma  \in \mathbb{R}$. The potential wells associated with these operators $V^{(\sigma)}(\overline{x})=- \sigma  (\sigma  +1) \, {\rm sech}^2 \overline{x}$ are depicted in Figure 1 where the integer values of $\sigma$ has been emphasized. Note that the wells become deeper and higher as $\sigma$ increases.

\begin{figure}[ht]
\centerline{
\includegraphics[height=3.5cm]{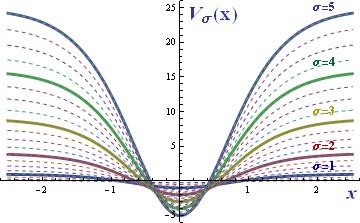} \hspace{2.5cm} \includegraphics[height=3.5cm]{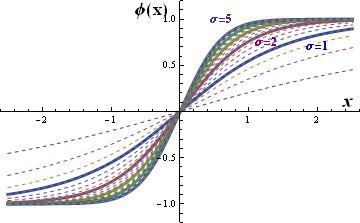} }
\caption{\small Graphical representations of: (a) the potential well in the Hessian operator for several values of $\sigma$ (b) the kink profiles for several values of $\sigma$. The cases $\sigma=N$ have been emphasized.}
\end{figure}
We shall denote by $\phi_K^{(\sigma)}(\overline{x})$ the kink solutions and by $U^{(\sigma)}(\phi)$ the parent potential  for a given $\sigma$ in this family of models. Because the kink solutions asymptotically reach the vacuum values $\phi^{(\sigma)}_V=\pm 1$, the kink Hessian operators are accompanied by the vacuum Hessians:
\[
K^{(\sigma)}_0=-\frac{d^2}{dx^2}+\sigma^2 \quad .
\]
Thus, the mass of the fundamental quanta in these models is: $\frac{\partial^2 U}{\partial \phi^2}[\phi_V]=\sigma^2$.

\subsection{Zero modes and kinks}
Our first task is the identification of the parent potentials such that:
\begin{equation}
\frac{d^2 U^{(\sigma)}}{d\phi^2} [\phi_K^{(\sigma)}(\overline{x})] = \sigma^2- \sigma  (\sigma  +1) \, {\rm sech}^2 \overline{x} \qquad .
\label{condition01}
\end{equation}
The solution of (\ref{condition01}) seems hopeless without knowledge of $\phi_K^{(\sigma)}(\overline{x})$. The equation (\ref{ode1})
\begin{equation}
\frac{d\phi_K^{(\sigma)}}{d\overline{x}}= \sqrt{2U^{(\sigma)}(\phi_K^{(\sigma)})}
\label{edo02}\qquad ,
\end{equation}
however, does allow us to interpret (\ref{condition01}) as an ODE in the $\overline{x}$ variable via the change of variables from $\phi$ to $\overline{x}$. Recall that $\phi_K^{(\sigma)}:\mathbb{R} \rightarrow [-1,1]$, $\overline{x} \mapsto \phi_K^{(\sigma)}(\overline{x})$ is a bijective map determined by equation {(\ref{edo02}). Trading the implicit parent potential for the kink solutions by the explicit function of $\overline{x}$ $
\mathfrak{U}_\sigma(\overline{x})= U^{(\sigma)}(\phi_K^{(\sigma)}(\overline{x}))$ and bearing in mind from equation (\ref{ode1}) that
$\frac{dU^{(\sigma)}}{d\phi}=\frac{1}{\sqrt{2\mathfrak{U}_\sigma}} \frac{d\mathfrak{U}_\sigma}{d\overline{x}}$ and $\frac{d^2 U^{(\sigma)}}{d\phi^2} = \frac{1}{2\mathfrak{U}_\sigma}\frac{d^2 \mathfrak{U}_\sigma}{d\overline{x}^2} - \frac{1}{(2\mathfrak{U}_\sigma)^2}(\frac{d\mathfrak{U}_\sigma}{d\overline{x}})^2$,
the equation (\ref{condition01}) becomes:
\begin{equation}
\frac{1}{2 \mathfrak{U}_\sigma(\overline{x})} \frac{d^2 \mathfrak{U}_\sigma}{d \overline{x}^2} - \frac{1}{(2 \mathfrak{U}_\sigma(\overline{x}))^2} \left(\frac{d \mathfrak{U}_\sigma}{d \overline{x}}\right)^2 =  \sigma^2- \sigma  (\sigma  +1) \, {\rm sech}^2 \overline{x} \qquad .
\label{condition02}
\end{equation}
(\ref{condition01}) is a non-linear ordinary differential equation that can easily be reduced to the Riccati equation
by means of the change $F(x)=\frac{1}{2}\log\mathfrak{U}_\sigma$:
\[
\frac{d^2 F}{d \overline{x}^2}+\left(\frac{dF}{d\overline{x}}\right)^2=\sigma^2- \sigma  (\sigma  +1) \, {\rm sech}^2 \overline{x} \quad .
\]
A particular solution can be obtained by the standard procedure of sending the Riccati equation to the Schr$\ddot{\rm o}$dinger equation via the new change $\frac{d F}{dx}=\frac{d \log \psi_0}{dx}$, or, $\mathfrak{U}=\psi_0^2$:
\begin{equation}
\frac{d^2\psi_0}{d \overline{x}^2}=\left(\sigma^2 -\sigma(\sigma+1) {\rm sech}^2\overline{x}\right)\psi_0 \label{zerm}\qquad ,
\end{equation}
i.e., $\psi_0(\overline{x})=\sqrt{C_\sigma}{\rm sech}^\sigma(\overline{x})$ is the ground state of zero energy
in the PT potential well. Therefore, we find a solution depending on the integration constant $C_\sigma $ of our problem:
\begin{equation}
\mathfrak{U}_\sigma(\overline{x}) = C_\sigma \, {\rm sech}^{2\sigma} \overline{x} \quad , \quad \lim_{\overline{x}\rightarrow \pm \infty} \mathfrak{U}_\sigma(\overline{x}) = 0 \qquad .
\label{potentialinx}
\end{equation}
This solution fits smoothly into our scheme because the kinks tend to the zeroes of the potential $U^{(\sigma)}(\phi)$ at $x=\pm\infty$. The hidden reason for finding this expression (\ref{potentialinx}) for the parent potential as a function of the parameter $\overline{x}$ is that via the equation
\begin{equation}
d\phi_K^{(\sigma)} = \sqrt{2 \, \mathfrak{U}_\sigma(\overline{x})} d\overline{x}
\label{edo01}
\end{equation}
the potential is related to the translational mode $\frac{d\phi_K^{(\sigma)}}{d\overline{x}}$, which is in turn the ground state of zero energy of $K^{(\sigma)}$ because of general arguments (just plug it in (\ref{ode})). Integration of (\ref{edo01}) immediately produces the kink solution
\begin{equation}
\phi_K^{(\sigma)}(\overline{x}) = \frac{2\Gamma[\frac{1}{2}+\frac{\sigma}{2}]}{\sqrt{\pi} \Gamma[\frac{\sigma}{2}]} \tanh \overline{x} \,\, {}_2F_1[{\textstyle \frac{1}{2},1-\frac{\sigma}{2},\frac{3}{2},\tanh^2 \overline{x}}]
\label{kinkN1}
\end{equation}
in terms of a Gauss hypergeometric function. The constant $C_\sigma$ has been tuned to
\[
C_\sigma=\frac{2 \Gamma[\frac{1}{2}+\frac{\sigma}{2}]^2}{\pi \Gamma[\frac{\sigma}{2}]^2}
\]
in order to set the zeroes of the parent potential at the $\phi=\pm 1$ values. One can check that for $\sigma=1$ the soliton of the sine-Gordon model $\phi_K^{(1)}(\overline{x})= \frac{4}{\pi} \arctan \tanh \frac{\overline{x}}{2}$ is found, whereas the kink of the $\phi^4$ model  $\phi_K^{(2)}(\overline{x})=\tanh \overline{x}$ arises when $\sigma=2$.

It is convenient to rewrite the kink profile (\ref{kinkN1}) using other special functions. We recall the definition of the beta function in terms of three Gauss Gamma functions: $B(a,b)=\int_0^1 t^{a-1} (1-t)^{b-1}dt=\frac{\Gamma[a]\Gamma[b]}{\Gamma[a+b]}$. We also need the incomplete beta function $B(z;a,b)$ of a free upper integration limit $z\in [0,1]$: $B(z;a,b)=\int_0^x t^{a-1} (1-t)^{b-1}dt=\frac{1}{a} z^a \, {}_2F_1[a,1-b,a+1,z]$, which is related to the hypergeometric Gauss function. Finally, we shall consider the regularized incomplete beta function $I(z;a,b)=\frac{B(z;a,b)}{B(a,b)}$. With this arsenal we obtain the very compact formula:
\begin{equation}
\phi_K^{(\sigma)}(\overline{x})= {\rm sign}(\overline{x})\,\, I[\textstyle \tanh^2 \overline{x};\frac{1}{2},\frac{\sigma}{2}] \quad .
\label{kinkN2}
\end{equation}
Note that formula (\ref{kinkN2}) is an odd function because of the sign function that multiplies $I[\textstyle \tanh^2 \overline{x};\frac{1}{2},\frac{\sigma}{2}]$. This does not induce singularities because $I$ is zero at the origin, see Figure 2. It is remarkable that the soliton and kink solutions are embraced in this general formula.
\subsection{Recurrence relations}
The magic of the hypergeometric functions gives rise to interesting recurrence relations between kinks for different values of $\sigma$ from (\ref{kinkN1}). The well-known formula ${}_2F_1[a,b+1,c,z] ={}_2F_1[a,b,c,z]+\frac{z}{b} \frac{d}{dz} {}_2F_1[a,b,c,z]$ applied in (\ref{kinkN1}) means that:
\begin{equation}
\phi_K^{(\sigma+2)}(\overline{x}) = \phi_K^{(\sigma)}(\overline{x}) + \frac{1}{\sigma} \tanh \overline{x} \, \frac{d\phi_K^{(\sigma)}}{d\overline{x}}(\overline{x})
\label{recurrence01} \quad ,
\end{equation}
a relation between the $\sigma$ and $\sigma +2$ kink profiles. Taking into account (\ref{edo01}) we rewrite (\ref{recurrence01}) as:
\begin{equation}
\phi_K^{(\sigma+2)}(\overline{x}) = \phi_K^{(\sigma)}(\overline{x}) + \frac{2 \Gamma[\frac{1}{2}+\frac{\sigma}{2}]}{\sigma \sqrt{\pi}\Gamma[\frac{\sigma}{2}]} \tanh \overline{x} \, {\rm sech}^\sigma \overline{x} \quad ,
\label{recurrence02}
\end{equation}
 a form of the recurrence relations that can also be derived from the identities between regularized beta functions: $I(z,a,b)=1-I(1-z,b,a)$ and $I(z,a+1,b)=I(z,a,b)-\frac{z^a (1-z)^b}{aB(a,b)}$, applied in the formula (\ref{kinkN2}). In formula (\ref{recurrence02}) one notices that the $(\sigma+2)$-kink is equal to the addition of the $\sigma$-kink plus an odd function, implying that $|\phi_K^{(\sigma+2)}(\overline{x})|\geq |\phi_K^{(\sigma)}(\overline{x})|$. Therefore, the kink profile is more and more concentrated around its center in members of the family with higher and higher $\sigma$, see Figure 1(b).

For odd and even integer values of $\sigma$ the formulas simplify analytically. We discuss the $\sigma=2k$ even and $\sigma=2k+1$ odd, $k\in \mathbb{N}$, cases separately. If $\sigma=2k$, (\ref{recurrence02}) becomes:
\[
\phi_K^{(2k+2)}(\overline{x}) = \phi_K^{(2k)}(\overline{x}) + \frac{(2k-1)!!}{2^k k!} \tanh \overline{x} \, {\rm sech}^{2k} \overline{x} \quad .
\]
In the odd case the same recurrence relation reads:
\[
\phi_K^{(2k+3)}(\overline{x}) = \phi_K^{(2k+1)}(\overline{x}) + \frac{2^{k+1}(k+1)!}{\pi (2k+1)!!} \tanh \overline{x} \, {\rm sech}^{2k+1} \overline{x} \quad .
\]
In Table 1 we show the kink profiles of the first ten members of the hierarchy of models.

\begin{table}[ht]
\centerline{\begin{tabular}{|c|l|} \hline
$\sigma$ & $\phi_K^{(\sigma)}(\overline{x})$ \\ \hline
1 & $\phi_K^{(1)}(\overline{x})=\frac{2}{\pi} \arcsin \tanh \overline{x}$ \\
3 & $\phi_K^{(3)}(\overline{x})=\phi_K^{(1)}(\overline{x})+ \frac{2}{\pi} \tanh \overline{x}\, {\rm sech}\, \overline{x}$ \\
5 & $\phi_K^{(5)}(\overline{x})=\phi_K^{(3)}(\overline{x})+ \frac{4}{3\pi} \tanh \overline{x}\, {\rm sech}^3 \overline{x}$ \\
7& $\phi_K^{(7)}(\overline{x})=\phi_K^{(5)}(\overline{x})+ \frac{16}{15\pi} \tanh \overline{x}\, {\rm sech}^5 \overline{x}$ \\
9 & $\phi_K^{(9)}(\overline{x})=\phi_K^{(7)}(\overline{x})+ \frac{32}{35\pi} \tanh \overline{x}\, {\rm sech}^7 \overline{x}$ \\ \hline
\end{tabular}\hspace{0.5cm}\begin{tabular}{|c|l|} \hline
$\sigma$ & $\phi_K^{(\sigma)}(\overline{x})$ \\ \hline
2 & $\phi_K^{(2)}(\overline{x})=\tanh \overline{x}$ \\
4 & $\phi_K^{(4)}(\overline{x})=\phi_K^{(2)}(\overline{x})+ \frac{1}{2} \tanh \overline{x}\, {\rm sech}^2 \overline{x}$ \\
6 & $\phi_K^{(6)}(\overline{x})=\phi_K^{(4)}(\overline{x})+ \frac{3}{8} \tanh \overline{x}\, {\rm sech}^4 \overline{x}$ \\
8 & $\phi_K^{(8)}(\overline{x})=\phi_K^{(6)}(\overline{x})+ \frac{5}{16} \tanh \overline{x}\, {\rm sech}^6 \overline{x}$ \\
10 & $\phi_K^{({10})}(\overline{x})=\phi_K^{(8)}(\overline{x})+ \frac{35}{128} \tanh \overline{x}\, {\rm sech}^8 \overline{x}$ \\ \hline
\end{tabular}}
\caption{\small Some kink solutions of the hierarchy of models characterized by $\sigma\in\mathbb{N}$.}
\end{table}

 There also exist recurrence relations between $\sigma$- and $\sigma+1$-kinks differing in one unit in the family parameter, although in this case the formula involves a quadrature. From (\ref{edo01}) we obtain
\[
d\phi_K^{({\sigma+1})} = \frac{\sqrt{\mathfrak{U}_{\sigma+1}(\overline{x})}}{\sqrt{\mathfrak{U}_{\sigma}(\overline{x})}} d\phi_K^{(\sigma)}
=\frac{\sigma}{2} \frac{\Gamma[\frac{\sigma}{2}]^2}{\Gamma[\frac{1}{2}+\frac{\sigma}{2}]^2} \, {\rm sech}\, \overline{x}\, d\phi_K^{(\sigma)}\quad ,
\]
which can be two-sided integrated to find:
\[
\phi_K^{(\sigma+1)}(\overline{x}) = \frac{\sigma}{2}\frac{\Gamma[\frac{\sigma}{2}]^2}{\Gamma[\frac{1}{2}+\frac{\sigma}{2}]^2} \left[ {\rm sech}\, \overline{x}\, \phi_K^{(\sigma)} (\overline{x})+ \int {\rm sech}\, \overline{x}\, \tanh \overline{x} \, \phi_K^{(\sigma)}(\overline{x}) d\overline{x}\right] \quad .
\]
An integration by parts has been performed in the right-hand side member.

The classical energy of the kinks can easily be worked out from the zero modes, the connection formulas between hypergeometric functions and the Kummer theorem:
\[
E_{\rm clas}[\phi_K^{(\sigma)}(\overline{x})]=\int_{-\infty}^\infty\, d\overline{x} \,\left(\frac{d\phi_K^{(\sigma)}}{d\overline{x}}\right)^2=\frac{4^{\sigma +1}}{\pi\sigma}\cdot\frac{\Gamma[\frac{1}{2}+
\frac{\sigma}{2}]^2}{\Gamma[\sigma+\frac{1}{2}]^2}\cdot {}_2F_1[\sigma,2 \sigma,1+\sigma;-1] = \frac{4 \Gamma[\sigma] \Gamma[\frac{1}{2}+\frac{\sigma}{2}]^2}{\sqrt{\pi} \Gamma[\sigma+\frac{1}{2}] \Gamma[\frac{\sigma}{2}]^2}\quad .
\]
\subsection{Parent potentials}

To completely define the family of models we identify the potential (\ref{potentialinx}) as a function of the field $\phi$ by, first, inverting formula (\ref{kinkN2}):
\[
\overline{x} ={\rm arctanh}\, \sqrt{I^{-1}\textstyle \left[|\phi_K^{(\sigma)}|;\frac{1}{2},\frac{\sigma}{2}\right]}
\]
and, then,  plugging this expression into the formula (\ref{potentialinx}) to find:
\begin{equation}
U^{(\sigma)}(\phi)=\frac{2 \Gamma[\frac{1}{2}+\frac{\sigma}{2}]^2}{\pi \Gamma[\frac{\sigma}{2}]^2} \left(1- I^{-1}\textstyle \left[|\phi|;\frac{1}{2},\frac{\sigma}{2} \right]\right)^\sigma  \qquad .
\label{potentialinPhi}
\end{equation}
We emphasize that this formula is only valid in the interval $\phi\in[-1,1]$ in the internal space, see Figure 2(a). The reason is that the kink solutions live only in this interval and use of the map $\phi_K^{(\sigma)}:\mathbb{R}\rightarrow [-1,1]$ to reconstruct the potential in (\ref{action}) is restricted to this interval. Obviously the character of the potential for field configurations such that $|\phi|\geq 1$ does not influence the kink solutions. We have the freedom, however, to mathematically extend the definition of the potential as a function of $\phi$ to the $|\phi|\geq 1$ range. Even requiring twice-differentiable functions $U^{(\sigma)}(\phi)\in {\cal C}^2(\mathbb{R})$, there are several possible choices of continuations of $U^{(\sigma)}(\phi)$: see \cite{Trullinger1987} for some examples. In Figure 2(b) we show periodic continuations of the potential defined in $\phi\in[-1,1]$ to the $|\phi|\geq 1$ half-lines. Here, the piece of $U$ depicted in the Figure 2(a) is cloned in all the $[2k-1,2k+1]$, $k\in \mathbb{Z}$, intervals. The continuations depicted in Figure 2(c) are accomplished analytically by expanding $
U^{(\sigma)}(\phi)=\frac{2 \Gamma[\frac{1}{2}+\frac{\sigma}{2}]^2}{\pi \Gamma[\frac{\sigma}{2}]^2} \left(1- I^{-1}\textstyle \left[|\phi|;\frac{1}{2},\frac{\sigma}{2} \right]\right)^\sigma$ as a power series near the $\phi=\pm 1$ points and truncating the series to a polynomial at a finite higher than two order.

\begin{figure}[ht]
\centerline{
\includegraphics[height=3.cm]{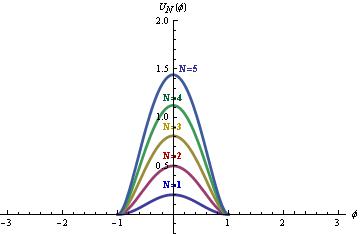} \hspace{1.5cm} \includegraphics[height=3.cm]{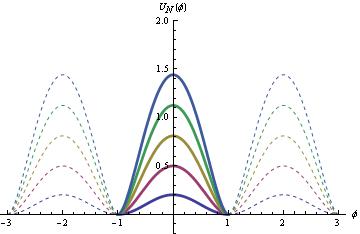} }
\end{figure}
\vspace{0.5cm}
\begin{figure}[ht]
\centerline{\includegraphics[height=3.cm]{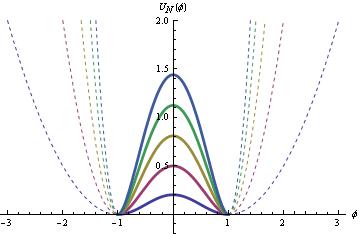} \hspace{1.5cm} \includegraphics[height=3.cm]{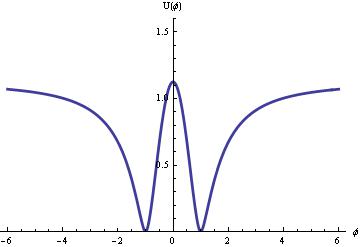} }
\caption{\small Graphical representation for the potential: (a) in the range $|\phi|\leq 1$, (b) periodic continuations, (c) polynomial continuations, (d) $\sigma=4$ generalized Chebychev continuation, to $|\phi|\geq 1$.}
\end{figure}

 Starting from the $\sigma=1$ potential in formula (\ref{potentialinPhi}) extended by the periodic continuation scheme we obtain: $U^{(1)}(\phi)=\frac{1}{\pi^2} [1+\cos(\pi \phi)]$, a beautiful version  of the sine-Gordon model (properly translated in internal space). Extending the $\sigma=2$ potential in (\ref{potentialinPhi}) by using the second type of continuation we recover the $\phi^4$ model: $U^{(2)}(\phi)=\frac{1}{2}(\phi^2-1)^2$. Indeed, these two examples are the exceptional ${\cal C}^\infty(\mathbb{R})$ continuations in the whole hierarchy of $\sigma=N$ potentials. More generally, $C^3(\mathbb{R})$ continuations of the potential (\ref{potentialinPhi}) are not possible if $\sigma> 2$ because
\[
\frac{d^3 U^{(\sigma)}}{d\phi^3} = \frac{d}{d\phi} \left[ \sigma^2 - \sigma(\sigma+1) \,{\rm sech}^2\, \overline{x} \right] = 2\sigma(\sigma+1)\, {\rm sech}^2\, \overline{x} \tanh \overline{x} \frac{d\overline{x}}{d\phi} = \frac{2\sigma(\sigma+1)}{\sqrt{C_\sigma}}\, {\rm sech}^{2-\sigma}\overline{x} \tanh \overline{x}
\]
diverges when $\overline{x}$ tends to infinity (the vacuum points $\phi=\pm 1$). There is however another case, $\sigma=4$, where the (\ref{potentialinPhi}) $U^{(4)}(\phi)= \frac{9}{8} \left[  1 - 2 \cos \left( \frac{2}{3} \arcsin \phi\right) \right]^4$ extended to the whole real line $\mathbb{R}$ can be expressed in terms of elementary functions:
\begin{equation}
U^{(4)}(\phi)= \left\{ \begin{array}{ll} \frac{9}{8} \left[  1 - 2 \cos \left( \frac{2}{3} \arcsin \phi\right) \right]^4 & \mbox{ if } |\phi|\leq 1 \\ \frac{9}{8} \left[  1 - 2 \cos \left( \frac{2}{3} \arcsin \frac{1}{\phi} \right) \right]^4 & \mbox{ if } |\phi|> 1 \end{array} \right. \quad .
\label{Potential04}
\end{equation}
This function, akin to Chebychev polynomials, is depicted in Figure 2(d) and is easily seen to be ${\cal C}^2(\mathbb{R})$. Note, however, that this continuation differs from that depicted in Figure 2(c), in particular the extended potential tends to $\frac{9}{8}$ at $\phi=\pm\infty$.

\subsection{Kink stability}
We recall that the Hessian differential operator at the vacuum solutions $\phi_V^{(\sigma)}=\pm 1$ reads:
\[
K_0^{(\sigma)}=-\frac{d^2}{dx^2} +\frac{\partial^2 U^{(\sigma)}}{\partial \phi^2}[\pm 1] = -\frac{d^2}{dx^2} + \sigma^2 \quad ,
\]
see (\ref{hessianoV}). In order to analyze the spectrum of this differential operator we temporarily confine the system in an interval of (non dimensional) length $l$ and impose periodic boundary conditions on the plane eigen-waves $\psi_{0k}^{(\sigma)}(x)=\frac{1}{\sqrt{l}}e^{-i k x}$. The wave vectors are quantized $k l=2\pi n, n\in{\mathbb Z}$, and in the limit of very large $l$ the spectral density is: $\rho(K_0^{(\sigma)})=\frac{d n}{d k}=\frac{l}{2\pi}$. The corresponding eigenvalues are $\omega^2(k)=k^2+\sigma^2$ such that the threshold of the continuous spectrum arising at $l=\infty$ is $\sigma^2$. The constant function is also a non-propagating
eigen-function of $K_0^{(\sigma)}$ with energy $\sigma^2$. Bound states buried in the threshold of the continuous spectrum are called half-bound states because due to the 1D Levinson Theorem must be counted with a weight of one-half. In sum, we have that ${\rm Spec}(K_0^{(\sigma)})=\{\sigma^2\}_{\frac{1}{2}}\cup \{k^2 +\sigma^2\}_{k\in \mathbb{R}}$. All the eigenvalues are positive, and there are no tachyonic waves fluctuating around the vacua, which accordingly are stable.

  The analysis of kink stability requires us to deal with a more complicated but tractable spectral problem: $K^{(\sigma)} \psi_\omega(x)= \omega^2 \psi_\omega(x)$. We confront the Schr\"odinger operator
\begin{equation}
K^{(\sigma)}=-\frac{d^2}{dx^2}+v^2 +V(x)=-\frac{d^2}{dx^2}+\sigma^2-\sigma(\sigma+1) \,{\rm sech}^2 x  \qquad ,
\label{hessiano}
\end{equation}
i.e., the quantum Hamiltonian describing the motion of a particle of (non-dimensional) mass $\mu=\frac{\hbar^2 \gamma_d^4}{2}$ in a $x\to -x$ symmetric P\"oschl-Teller well (\ref{PoschtTellerPotential}). The scattering eigen-functions of eigenvalue $\omega^2(k;\sigma)=k^2+\sigma^2$ are well known \cite{DJ}:
\begin{equation}
\psi_k^{(\sigma)}(x)=2^{i k} C_k^{(\sigma)} {\rm sech}^{-i k}\overline{x}\, \, \cdot {}_2F_1 \left[\sigma+1-ik,-\sigma -ik,1-i k,{\textstyle\frac{e^{-\overline{x}}}{e^{\overline{x}}+e^{-\overline{x}}}}\right] \hspace{1cm} k\in \mathbb{R}
\label{continuousstates}\quad .
\end{equation}
These eigen-functions belonging to the continuous spectrum of $K^{(\sigma)}$ can also be written in the simpler form
\begin{equation}
\psi_k^{(\sigma)}(x)= C_k^{(\sigma)} e^{i k \overline{x}} {}_2F_1 \left[-\sigma,\sigma+1,1-i k,{\textstyle\frac{e^{-\overline{x}}}{e^{\overline{x}}+e^{-\overline{x}}}}\right]
\label{continuousstates1}
\end{equation}
by using the linear transformation: ${}_2F_1[a,b,c,z]=(1-z)^{c-a-b}{}_2F_1[c-a,c-b,c,z]$. From the asymptotic behaviour
\begin{eqnarray*}
\psi_k^{(\sigma)}(x)&&\simeq_{x\to\infty} C_k^{(\sigma)} e^{i k \overline{x}}\\
\psi_k^{(\sigma)}(x)&&\simeq_{x\to -\infty} C_k^{(\sigma)}\left\{\frac{\Gamma(1-ik)\Gamma(-ik)}{\Gamma(\sigma +1 -ik)\Gamma(-\sigma-ik)}e^{i k \overline{x}}+\frac{\Gamma(1-ik)\Gamma(ik)}{\Gamma(\sigma +1)\Gamma(-\sigma)}e^{i k \overline{x}}e^{-i k \overline{x}}\right\} \quad,
\end{eqnarray*}
read in (\ref{continuousstates1}), the transmission and reflection amplitudes are identified:
\begin{equation}
t^{(\sigma)}(k)= \frac{\Gamma(\sigma +1-ik)\Gamma(-\sigma -ik)}{\Gamma(1-ik)\Gamma(-ik)}\, \, \, , \quad
r^{(\sigma)}(k)= \frac{\Gamma(\sigma +1-ik)\Gamma(-\sigma -ik)\Gamma(ik)}{\Gamma(1+\sigma)\Gamma(-\sigma)\Gamma(-ik)}\quad . \label{asdata}
\end{equation}
The phase shifts in the even, $\delta_+^{(\sigma)}(k)$, and odd, $\delta_-^{(\sigma)}(k)$, channels are related to these amplitudes via the identities:
\begin{eqnarray*}
e^{2i\delta_\pm^{(\sigma)}}(k)&=& t^{(\sigma)}\pm r^{(\sigma)}(k) \, \, \, , \qquad \delta_\pm^{(\sigma)}(k)=\frac{1}{2}{\rm arctan}\left[\frac{{\rm Im}(t^{(\sigma)}\pm r^{(\sigma)}(k))}{{\rm Re}(t^{(\sigma)}\pm r^{(\sigma)}(k))}\right]\\ t^{(\sigma)}(k)&=& {\rm cos}\left(\delta_+^{(\sigma)}(k)-\delta_-^{(\sigma)}(k)\right)\cdot {\rm exp}[\frac{i}{2}(\delta_+^{(\sigma)}(k)+\delta_-^{(\sigma)}(k))] \\  r^{(\sigma)}(k)&=& \sin \left(\delta_+^{(\sigma)}(k)-\delta_-^{(\sigma)}(k)\right)\cdot {\rm exp}[\frac{i}{2}(\delta_+^{(\sigma)}(k)+\delta_-^{(\sigma)}(k)+\pi)]\quad .
\end{eqnarray*}
Again we impose PBC $e^{ikx}=e^{ik(x+l)}\cdot e^{\frac{i}{2}(\delta_+^{(\sigma)}(k)+\delta_-^{(\sigma)}(k))}$
such that $kl+ \frac{1}{2}(\delta_+^{(\sigma)}(k)+\delta_-^{(\sigma)}(k))=2\pi n$ and the spectral density of the
kink fluctuations is, for very large $l$: $\rho(K^{(\sigma)})=\frac{d n}{d k}=\frac{l}{2\pi} + \frac{1}{2\pi} \frac{d \delta^{(\sigma)}}{d k}(k)$, where $\delta^{(\sigma)}(k)=\delta^{(\sigma)}_+(k)+\delta^{(\sigma)}_-(k)$ are the total phase shifts -the sums of the phase shifts in the even and odd channels- of the scattering eigenfunctions of $K^{(\sigma)}$.

Strictly at the $l=\infty$ limit the spectrum of $K^{(\sigma)}$ includes a discrete part. There are several bound states: $k=j(\sigma -j), j=0,1,2, \cdots , I[\sigma],$ are $I[\sigma]+1$ poles of $t^{(\sigma)}(k)$ in the positive imaginary axis of the complex $k$-plane. Here, $I[\sigma]$ denotes the integer part of $\sigma$ and we temporarily  assume that $I[\sigma]\neq\sigma$. $ \omega_j^2(\sigma)=-(\sigma-j)^2+\sigma^2=j(2\sigma -j)$ are the eigenvalues of these $I[\sigma]+1$ bound states, whereas the eigenfunctions, derived from (\ref{continuousstates}) for a better visualization of the $L^2$-integrability character, read:
\[
\psi_j^{(\sigma)}(x) =\frac{C_j^{(\sigma)}}{(e^x+e^{-x})^{\sigma-j}} \, {}_2 F_1 \textstyle [2\sigma+1-j,-j,\sigma+1-j,\frac{e^{-x}}{e^x+e^{-x}}]\quad .
\]
Note that
$\omega_0^2(\sigma)=0, \forall\sigma$. The ever present zero-mode eigen-function is the Goldstone boson owing
to the spontaneous breaking of the translational symmetry by the kink.

 In the special cases when $\sigma=N\in \mathbb{N}$ is a natural number two related novelties arise:
 \begin{enumerate}
\item The reflection amplitude $r^{N}(k)=0$ becomes null and the kink potential well is transparent to the
motion of mesons. The transition amplitude, however, becomes:
\begin{equation}
t^{(N)}(k)=-\Pi_{j=0}^{N-1}\, \frac{N-j-ik}{N-j+ik}=\Pi_{j=0}^{N-1}\, \frac{k^2-(N-j)^2+2ik(N-j)}{(N-j)^2+k^2}\label{trana} \quad .
\end{equation}
 \item There are only $N$ bound states in the spectrum of $K^{(N)}$, although a half-bound state just buried at the threshold $N^2$ of the continuous spectrum arises:
\[
\psi_\frac{1}{2}^{(N)}(\overline{x}) = C_\frac{1}{2}^{(\sigma)} \, {}_2 F_1 \textstyle [N+1, -N,1,\frac{e^{-x}}{e^x+e^{-x}}] \quad .
\]
\end{enumerate}

In sum, the spectrum of $K^{(\sigma)}$ is slightly different in two situations: (1) if $\sigma=N\in \mathbb{N}$ we have: ${\rm Spec}(K^{(N)})=\{j(2N-j)\}_{j=0,1,\dots, N-1}\cup \{N^2\}_\frac{1}{2}\cup \{k^2 +N^2\}_{k\in \mathbb{R}}$. (2) if $\sigma\notin \mathbb{N}$, ${\rm Spec}(K^{(\sigma)})=\{j(2\sigma-j)\}_{j=0,1,\dots, I[\sigma]}\cup \{k^2 +\sigma^2\}_{k\in \mathbb{R}}$. Therefore, the spectrum of the Hessian operator $K^{(\sigma)}$ is non-negative and the kink solutions are stable in all these models.

\section{One-loop quantum correction to the kink masses}

In this Section we address the main issue in this work: computation of the quantum correction to the classical kink masses for arbitrary values of $\sigma\in \mathbb{R}^+$. We stress that as far as one-loop (Gaussians) fluctuations are concerned this is a well defined problem, see \cite{Christ1975,Trullinger1987, Boya1989}. It seems to us convenient to distinguish the $\sigma=N\in \mathbb{N}$ and $\sigma\notin \mathbb{N}$ cases in these computations. For the sake of completeness and to gain better knowledge of this subtle quantum phenomenon we shall address this task using different tools: the CCG formula, the DHN formula, the generalized zeta function regularization method, and the asymptotic heat kernel approach. The four methods lead to compatible results, reinforcing the reliability of each individual method and the conclusions obtained.

The one-loop kink mass -the classical kink mass plus the one-loop quantum correction- is of the form:
\[
\tilde{E}[\psi_K]_{\rm 1-loop}=\frac{m_d}{\gamma_d^2}\left(E[\phi_K]+\hbar\gamma_d^2\bigtriangleup E[\phi_k]\right)\quad .
\]
Thus, formally, the object to be computed is:
\begin{equation}
\bigtriangleup E[\phi_K^{(v)}]=-\frac{1}{4}{\rm Tr}_{L^2}\Big[\frac{(K^{\frac{1}{2}}-K_0^{\frac{1}{2}})^2}{K_0^{\frac{1}{2}}}\Big]=\frac{1}{2}
{\rm Tr}_{L^2}\Big[K^{\frac{1}{2}}-K_0^{\frac{1}{2}}-\frac{V(x)}{2 K_0^{\frac{1}{2}}}\Big]
\label{olms}
\end{equation}
where the operators entering formula (\ref{olms}) are:
\begin{equation}
K_0=\left. -\frac{d^2}{dx^2}+\frac{\partial^2 U}{\partial\phi^2}\right|_{\phi_V}=-\frac{d^2}{dx^2}+v^2 \, \, , \quad K=\left. -\frac{d^2}{dx^2}+\frac{\partial^2 U}{\partial\phi^2}\right|_{\phi_K}=-\frac{d^2}{dx^2}+v^2+V(x) \, \, \, .
\label{olflu}
\end{equation}
We must make sense of the $L^2$-trace of quotients and differences of square roots of differential operators to account for the effect of the kink quantum fluctuations measured with respect to the vacuum quantum fluctuations. In order to do so it is convenient to divide  $\bigtriangleup E[\phi_K^{(v)}]=\bigtriangleup E_1[\phi_K^{(v)}]+\bigtriangleup E_2[\phi_K^{(v)}]$ into two parts with different physical origins:
\begin{enumerate}
\item $\bigtriangleup E_1[\phi_K^{(v)}]=\frac{1}{2}
{\rm Tr}_{L^2}(K^{\frac{1}{2}}-K_0^{\frac{1}{2}})$ is due to the zero-point renormalization of the kink vacuum energy by subtracting the vacuum energy. We emphasize that this formula means that the subtraction is performed mode-by-mode. In infinite dimensional spaces $
{\rm Tr}_{L^2}(K^{\frac{1}{2}}-K_0^{\frac{1}{2}})\neq {\rm Tr}_{L^2}K^{\frac{1}{2}}-{\rm Tr}_{L^2} K_0^{\frac{1}{2}}$; the difference in the traces (not the trace of the difference) would arise if the total energy of the vacuum fluctuations were subtracted from the total energy of the kink vacuum fluctuations. This is a source of many controversies in the Literature.

\item $\bigtriangleup E_2[\phi_K^{(v)}]=-\frac{1}{4}
{\rm Tr}_{L^2}\Big(\frac{V(x)}{K_0^{\frac{1}{2}}}\Big)$ comes from the one-loop counter-term induced by the mass renormalization procedure in $(1+1)D$ scalar field theory merely due to normal-ordering of the Hamiltonian.

\end{enumerate}

\subsection{The $\sigma=N\in \mathbb{N}$ case}

\subsubsection{The Cahill-Comtet-Glauber (CCG) formula}
We shall always deal with potentials such that: $\int_{-\infty}^\infty\, dx \, (1+|x|)V(x)<+\infty$. If $\sigma =N\in \mathbb{N}$, the reflection amplitude is null, $r^{(N)}(k)=0$ and the potential well is transparent. Given these two circumstances, it is shown in Reference \cite{Cahill1976}, relying on Zakharov and Faddeev, \cite{ZF}, that:
\begin{enumerate}
\item There is a nice relation between the bound state eigenvalues and the phase shifts given by the formula:
\begin{equation}
\delta^{(v)}(k)= 2 \sum_j\, {\rm arctan}\Big[\frac{v}{k}\sin \theta_j\Big] \, \, , \quad \theta_j={\rm arccos}\frac{\omega_j}{v} \label{psbs} \quad .
\end{equation}
Here, $\omega_j^2$ are the bound state eigenvalues of $K$, i.e., those below the threshold of the continuous
spectrum: $0\leq \omega_j^2 < v^2$. The sum is, of course, over all these eigenvalues.

\item The total area enclosed by $V(x)$ is also related to the bound state eigenvalues:
\begin{equation}
\int_{-\infty}^\infty\, dx \, V(x)=-4 v\sum_j \, \sin \theta_j \label{abs} \qquad .
\end{equation}

\item There is a last feature in the spectrum of these reflectionless potentials to be taken into account: the highest bound state of $K$ sits exactly at the threshold of the continuous spectrum. The contribution of these eigen-functions -half-bound states- cancel exactly with the contribution of the constant eigenfunctions  -also half-bound states- of $K_0$.
\end{enumerate}
With all this information at our disposal, we derive the CCG formula. The first contribution $\bigtriangleup E_1[\phi_K^{(v)}]$ is
\begin{eqnarray*}
\hspace{-3cm}&&\frac{1}{2}
{\rm Tr}_{L^2}(K^{\frac{1}{2}}-K_0^{\frac{1}{2}})=\frac{1}{2}\sum_j \, \omega_j -\frac{1}{4\pi}\int_{-\infty}^\infty \, dk \, \delta^{(v)}(k)\frac{k}{\sqrt{k^2+v^2}}=\frac{v}{2}\sum_j\, \cos \theta_j-\\ && -\frac{1}{2\pi}\lim_{\Lambda\to\infty}\sum_j\, \left\{\left.\left(\sqrt{k^2+v^2}{\rm arctan}[\frac{v}{k}\sin \theta_j]\right)\right|_{-\Lambda}^\Lambda -\int_{-\Lambda}^\Lambda\, dk\, \sqrt{k^2+v^2}\frac{d}{dk}\left({\rm arctan}[\frac{v}{k}\sin \theta_j]\right)\right\}\\ &&= \frac{v}{\pi} \sum_j \left[\, \left(\frac{\pi}{2}-{\rm arctan}[{\rm cot}\theta_j]\right)\cos \theta_j- \sin \theta_j\right]-\frac{v}{2\pi}\sum_j\,\sin \theta_j \lim_{\Lambda\to\infty}\log\left[\frac{\Lambda+\sqrt{\Lambda^2+v^2}}{-\Lambda+\sqrt{\Lambda^2+v^2}}\right]\quad .
\end{eqnarray*}
while the second $\bigtriangleup E_2[\phi_K^{(v)}]$ reads
\begin{eqnarray*}
-\frac{1}{4}{\rm Tr}_{L^2}\Big(\frac{V(x)}{K_0^{\frac{1}{2}}}\Big)&=&-\frac{1}{8\pi}\int_{-\infty}^\infty \, dx \, V(x)\,\int_{-\infty}^\infty \, \frac{dk}{\sqrt{k^2+v^2}}=\frac{v}{2\pi}\sum_j \, \sin \theta_j \int_{-\infty}^\infty \, \frac{dk}{\sqrt{k^2+v^2}}\\ &=& \frac{v}{2\pi}\sum_j \, \sin \theta_j \lim_{\Lambda\to\infty}\log\left[\frac{\Lambda+\sqrt{\Lambda^2+v^2}}{-\Lambda+\sqrt{\Lambda^2+v^2}}\right]\quad .
\end{eqnarray*}
Adding these two pieces together the logarithmic divergences cancel and we obtain the very compact formula
\begin{equation}
\Delta E[\phi_K^{(v)}] = - \frac{v}{\pi}  \sum_j (\sin \theta_j - \theta_j \cos \theta_j)
\label{cahill}
\end{equation}
for the one-loop kink mass shift derived in \cite{Cahill1976} in terms of the threshold $v$ and the bound state eigenvalues $\omega_j^2$.

In the models studied in this paper the operator $K^{(N)}$ is of transparent P\"oschl-Teller type, supporting $N$ bound states with CCG angles:
\[
\theta_j=\arccos \frac{\sqrt{j(2N-j)}}{N} \qquad .
\]
Recall that $\omega_j^2(N)=j(2N-j)$ and $v^2=N^2$. Therefore, formula (\ref{cahill}), after some manipulations, provides the exact result in the form:
\begin{eqnarray}
\Delta E[\phi_K^{(N)}] & =& -\frac{N}{\pi} \sum_{j=0}^{N-1} \left[ \sin \arccos \frac{\sqrt{j(2N-j)}}{N} - \frac{\sqrt{j(2N-j)}}{N} \arccos \frac{\sqrt{j(2N-j)}}{N} \right]\nonumber\\ &=& -\frac{1}{2\pi} N(N+1)+\frac{1}{\pi} \sum_{r=1}^{N-1} \sqrt{N^2-r^2} \arcsin \frac{r}{N}\label{correction}\quad ,
\end{eqnarray}
where we have defined $r=N-j$.

The very well known results for the sine-Gordon- and $\phi^4$ kinks are reproduced here. If $N=1$, we find the one-loop quantum correction to the sG-soliton mass: $\Delta E[\phi_K^{(1)}]= -\frac{1}{\pi}$. For $N=2$, we obtain the one-loop quantum correction to the $\phi^4$-kink mass: $\Delta E[\phi_K^{(2)}]=  \frac{1}{2\sqrt{3}}
-\frac{3}{\pi}$ (in this case $m_d=\frac{m}{\sqrt{2}}$, but in general $m_d=\frac{m}{\sqrt{N}}$). In
Table 2 we display the results obtained by means of (\ref{correction}) from $N=1$ to $N=20$. A graphical representation of the dependence of the one-loop mass shift in $N$ is also depicted in the next Figure.

\begin{table}[ht]
\hspace{1.5cm}\begin{tabular}{|c|c|}
\hline  $N$ & $\Delta E(\phi_K^{(N)})$ \\ \hline
$1$ & $-0.31831$ \\
$2$ & $-0.66625$ \\
$3$ & $-1.08451$ \\
$4$ & $-1.58003$ \\
$5$ & $-2.15555$ \\
$6$ & $-2.81245$ \\
$7$ & $-3.55156$ \\
$8$ & $-4.37342$ \\
$9$ & $-5.27839$ \\
$10$ & $-6.26675$ \\ \hline
\end{tabular}
\hspace{0.6cm}
\begin{tabular}{|c|c|}
\hline  $N$ & $\Delta E(\phi_K^{(N)})$ \\ \hline
$11$ & $-7.33870$ \\
$12$ & $-8.49439$ \\
$13$ & $-9.73396$ \\
$14$ & $-11.0575$ \\
$15$ & $-12.4651$ \\
$16$ & $-13.9568$ \\
$17$ & $-15.5328$ \\
$18$ & $-17.1929$ \\
$19$ & $-18.9374$ \\
$20$ & $-20.7661$ \\ \hline
\end{tabular}
\hspace{0.6cm}
\begin{tabular}{c}
\includegraphics[height=4cm]{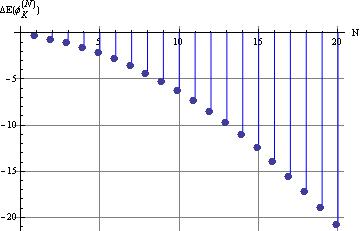}
\end{tabular}
\caption{\small Numerical values of the one-loop kink mass quantum corrections (\ref{kinkN2}), $\Delta E[\phi_K^{(N)}]$ for $N=1,\dots,20$ and graphical representation of the one-loop kink mass shift as a function of $N$.}
\end{table}

\subsubsection{The DHN formula}

In the seminal series of papers \cite{Dashen1974,Dashen1975,Dashen1975b} Dashen, Hasslacher and Neveu computed the one-loop quantum correction to the kink and soliton masses by using what is known as the \lq\lq first" DHN formula {\footnote{The \lq\lq second" DHN formula offered WKB computations on the energies of quantum sine-Gordon breathers.}}. Recently, the DHN prescription has been thoroughly revised by several research groups in order to address similar issues about supersymmetric kinks \cite{Rebhan1997,Goldhaber2004,Shifman1999,Graham1999,Bordag2002}. In this framework several subtleties arise, mainly, but not only, concerned with compatibilities between boundary conditions, regularization procedures, and supersymmetry. Like the CCG formula, which is no more than an elegant reformulation of the DHN result in terms of the bound state eigenvalues and the threshold of the continuous spectrum, the DHN procedure works fine when the potential wells involved are transparent. In this case, the DHN expresses the same $L^2$-trace combinations without using the connection between phase shifts and bound state eigenvalues encoded in the CCG formula:
\begin{eqnarray}
\Delta E[\phi_K]&=& \frac{1}{2}
{\rm Tr}_{L^2}(K^{\frac{1}{2}}-K_0^{\frac{1}{2}})-{\rm Tr}_{L^2}\frac{V(x)}{4 K_0^{\frac{1}{2}}} = \label{dhn} \\&=& \frac{1}{2}\sum_{j=0}^{N-1} \omega_j + \frac{1}{2\pi} \int_0^\infty dk \frac{d \delta}{dk} \sqrt{k^2 + v^2} + \frac{1}{4\pi} \left<V(x) \right>  -\frac{1}{8\pi}  \left<V(x) \right>  \int_{-\infty}^\infty \frac{dk}{\sqrt{k^2+v^2}} \nonumber\quad .
\end{eqnarray}
Because the kink P\"oschl-Teller wells are transparent when $\sigma=N$, the phase shifts $\delta^{(N)}(k)$ are only identified from the transmission amplitudes $t^{(N)}(k)$ (\ref{trana}) and, in turn, the kink spectral density
is easily derived:
\[
\rho[K^{(N)}](k)=\frac{l}{2\pi}+\frac{1}{2\pi}\frac{d\delta^{(N)}}{dk}=\frac{l}{2\pi} - \frac{1}{\pi} \sum_{r=1}^N \frac{r}{r^2+k^2} \qquad .
\]
The area of the kink well $V^{(N)}(x)=-N(N+1) \, {\rm sech}^2 \overline{x}$ is also easy to compute: $\left<V^{(N)}(x)\right>=-2N(N+1)$. Use of the identity $\frac{N(N+1)}{2}=\sum_{r=1}^N\, r$ and cancelation of the
$r=N$ terms allow the calculation of the integrals involved in formula (\ref{dhn}):
\begin{eqnarray*}
&& \hspace{-0.7cm}\frac{1}{2\pi} \int_0^\infty dk  \frac{1}{2\pi}\frac{d\delta^{(N)}}{dk}\sqrt{k^2 + N^2} -\frac{1}{8\pi}  \left<V(x) \right>  \int_{-\infty}^\infty \frac{dk}{\sqrt{k^2+N^2}} = \\ && =
\frac{1}{\pi} \lim_{\Lambda\rightarrow \infty}\int_0^\Lambda dk \left\{\left[- \sum_{r=1}^N \frac{r}{k^2+ r^2}  \right] \sqrt{k^2+N^2} + \frac{N(N+1)}{2\sqrt{k^2+N^2}}\right\} =
\\ && = \lim_{\Lambda\rightarrow \infty}\int_0^\Lambda dk \sum_{r=1}^{N-1}\frac{r}{\pi} \left[-\frac{\sqrt{k^2+N^2}}{k^2+ r^2}  + \frac{1}{\sqrt{k^2+N^2}} \right] =\\ && = \sum_{r=1}^{N-1} \frac{r(r^2-N^2)}{\pi} \lim_{\Lambda\rightarrow \infty}\int_0^\Lambda   \frac{dk}{(k^2+r^2)\sqrt{k^2+N^2}} = - \sum_{r=1}^{N-1} \frac{\sqrt{N^2-r^2}}{\pi} \, {\rm arccos}\, \frac{r}{N} \quad .
\end{eqnarray*}
Therefore, after adding the bound state contribution and subtracting the boundary term arising in the mode-by mode subtraction, in the DHN approach the one-loop quantum correction to the kink mass reads:
\begin{eqnarray*}
\Delta E[\phi_K^{(N)}]&=&\frac{1}{2} \sum_{j=1}^{N-1} \sqrt{j(2N-j)} - \frac{1}{2\pi} N(N+1) - \sum_{r=1}^{N-1} \frac{\sqrt{N^2-r^2}}{\pi} \, {\rm arccos}\, \frac{r}{N} =\\
&=&- \frac{1}{2\pi} N(N+1) + \frac{1}{\pi} \sum_{r=1}^{N-1} \sqrt{N^2-r^2} \left[ \frac{\pi}{2} - \, {\rm arccos}\, \frac{r}{N} \right] =\\
&=&- \frac{1}{2\pi} N(N+1) + \frac{1}{\pi} \sum_{r=1}^{N-1} \sqrt{N^2-r^2}  \arcsin \frac{r}{N} \quad .
\end{eqnarray*}
We have now performed the inverse change: $j=N-r$ to find an identical result to that offered by the CCG formula (\ref{correction}).

\subsubsection{The spectral zeta function procedure}

Traces and determinants of powers of elliptic operators are only defined by means of a process of analytic continuation that mimics the definition of the Riemann zeta function as a meromorphic function, giving formal meaning to strictly divergent series by extending the usually rational exponents of the eigenvalues to the complex plane. Replacing natural numbers by the eigenvalues of positive elliptic differential operators, the spectral or generalized zeta functions are defined. Assuming that $A\in {\cal A}[\overline{L}^2(\mathbb{R})]$ is an operator with a positive discrete spectrum, ${\rm Spec}\,A=\{\omega_n^2 \in \mathbb{R}: A\psi_n=\omega_n^2 \psi_n\}$, the spectral $A$-zeta function is the formal series obtained by summing over the $-s$-power of the eigenvalues:
\begin{equation}
\zeta_A(s)={\rm Tr}\,A^{-s} =\sum_{n} \frac{1}{\omega_n^{2s}} \hspace{0.5cm}, \hspace{0.5cm} s\in \mathbb{C}\quad .
\label{zetafunction}
\end{equation}
There is a general theory of elliptic pseudo-differential operators that characterizes the conditions under which the generalized zeta function is a meromorphic function. By a process of analytic continuation, close in spirit to the
analytic continuation defining the Riemann zeta function, the values that are not at the poles of the spectral zeta functions and their derivatives are taken as \lq\lq regularized" definitions of traces and logarithms of determinants of (complex powers of pseudo-)differential operators \cite{Gilkey1984, Roe1988, Vassilevich2003}.

We closely follow this idea to regularize the kink Casimir energy by using the spectral zeta functions of the $K$ and $K_0$ operators:
\begin{equation}
\Delta E_1(\phi_K)[s] = \frac{1}{2}
{\rm Tr}_{L^2}\left(K^\bot\right)^{-s}-{\rm Tr}_{L^2}K_0^{-s}=\frac{1}{2}\left(\sum_{n=1}^\infty\, \omega(n)^{-2s}-\sum_{n=1}^\infty\, \omega_0(n)^{-2s}\right)=\frac{1}{2}  [\zeta_{K^\bot}(s)-\zeta_{K_0}(s)]
\label{zetaregularization}
\end{equation}
appart from the physical pole: $s=-\frac{1}{2}$, \cite{Alonso2002}. By $K^\bot$ we denote the differential operator $K$ acting in the sub-space of the Hilbert space orthogonal to its kernel. From a mathematical point of view, zero modes are disastrous in the definition of spectral zeta functions, giving rise to uncontrollable divergences. Fortunately,  physical reasons show that zero modes due to the symmetry breaking of some continuous group only enter in two- and higher- orders in the loop expansion, see e.g. \cite{Rajaraman1982}. Simili modo, the other contribution coming from normal-ordered Hamiltonians can be also regularized:
\[
\Delta E_2(\phi_K)[s] = -\frac{1}{4}{\rm Tr}_{L^2}\left[V(x)K_0^{-s-1}\right]= \frac{1}{2} \left< V(x) \right>  \lim_{l\rightarrow \infty} \frac{1}{l} \frac{\Gamma[s+1]}{\Gamma[s]} \zeta_{K_0}(s+1) \quad .
\]
There is a very subtle issue here. By choosing the regularization point in $s+1$ instead of $s$ we pick the physical pole of both divergent quantities at the same point: $s=-\frac{1}{2}$ in ${\mathbb C}$. A miracle happens: the different residua provide, after cancelation of the divergent parts, exactly the same correction as the mode-by-mode subtraction, not present in the (\ref{zetaregularization}) term. In sum, the one-loop kink mass shift is obtained by this procedure through the formula:
\[
\Delta E(\phi_K)  =  \lim_{s\rightarrow -\frac{1}{2}} \left[ \Delta E_1(\phi_K)[s]  + \Delta E_2(\phi_K)[s]  \right] \qquad .
\]

It is also convenient to use another spectral function, the $K$-heat or partition function, because as a series in decaying exponentials of the eigenvalues this function enjoys better convergence properties and is related to the spectral zeta function via Mellin's transform:
\[
h_K(\beta)={\rm Tr}_{L^2}\, e^{-\beta K}=\sum_{n=1}^\infty\, e^{-\beta \omega^2(n)}
\quad , \quad \zeta_K(s)= \frac{1}{\Gamma[s]}\int_0^\infty d\beta \beta^{s-1} h_K(\beta) \quad ,
\]
where $\beta$ is a fictitious inverse temperature. In our models the $K$-heat function can be identified analytically  from the spectral data of the kink well:
\begin{eqnarray*}
{\rm Tr}_{L^2} e^{-\beta K^{(N)}}&=&\sum_{j=0}^N \, e^{-\beta j(2
N-j)}+\frac{l}{2\pi}\int_{-\infty}^\infty \, dk \, e^{-\beta
(k^2+N^2)}\left[1-\frac{2}{l}\sum_{r=1}^N \,
\frac{r}{r^2+k^2}\right]
\\ &=& e^{-\beta
N^2}\left(\frac{l}{\sqrt{4\pi\beta}}+\sum_{r=1}^N
\, e^{\beta r^2}{\rm Erf}[r\sqrt{\beta}]\right)
\end{eqnarray*}
 as a sum of error functions ${\rm Erf}[z]$. Because of the zero point vacuum renormalization and the exclusion of the zero mode the important quantity in the kink mass shift problem is:
\begin{equation}
h_{K^{(N)\perp}}(\beta) - h_{K_0^{(N)}}(\beta)={\rm Tr}_{L^2} e^{-\beta K^{(N)^\bot}}-{\rm Tr}_{L^2} e^{-\beta K_0^{(N)}} = e^{-N^2 \beta} \sum_{r=1}^{N} e^{r^2 \beta}\, {\rm Erf}\, (r\sqrt{\beta})-1 \quad ,
\label{Heat0}
\end{equation}
a magnitude that is plotted in Figure 3 as a function of $\beta$.
\begin{figure}[ht]
\centerline{\includegraphics[height=3cm]{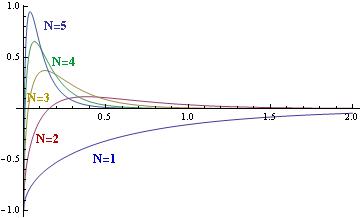}\hspace{0.4cm} \includegraphics[height=3cm]{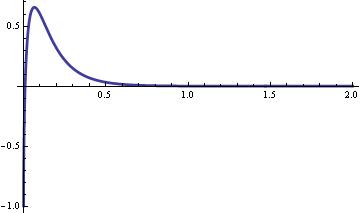} \hspace{0.4cm} \includegraphics[height=3cm]{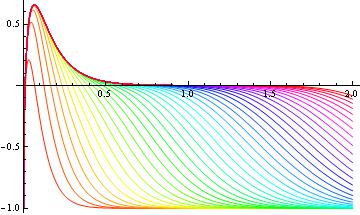} }
\caption{\small Graphical representation of the difference between the $K^{(N)^\bot}$-heat and $K_0^{(N)}$-heat functions: (a) the exact function for $N=1,2,3,4,5$, (b) the exact function for $N=4$, (c) the asymptotic
approximation for several $N_t$ truncation orders. Increasing values of $N_t$ give better approximations to the exact function.}
\end{figure}
The Mellin transform of formula (\ref{Heat0}) reads:
\begin{eqnarray*}
\zeta_{K^{(N)\perp}}(s)-\zeta_{K_0}(s) &=& \sum_{r=1}^{N-1} \left[ (N^2-r^2)^{-s} - \frac{1}{\sqrt{\pi} r^{2s}} \frac{\Gamma[s+\frac{1}{2}]}{\Gamma[s+1]} \, {}_2F_1[s,s+{\textstyle\frac{1}{2}},s+1,1-\textstyle{\frac{N^2}{r^2}}] \right] - \frac{\Gamma[s+\frac{1}{2}]}{\sqrt{\pi} N^{2s} s \Gamma[s]}
\\ &=& \sum_{r=1}^{N-1} \left[ \frac{2}{\sqrt{\pi} r^{2s}} \left( \frac{r^2}{N^2-r^2} \right)^{\frac{1}{2}+s} \frac{\Gamma[s+\frac{1}{2}]}{\Gamma[s]} \, {}_2F_1[{\textstyle\frac{1}{2}},s+{\textstyle\frac{1}{2}},{\textstyle\frac{3}{2}},-\textstyle{\frac{r^2}{N^2-r^2}}] \right] - \frac{\Gamma[s+\frac{1}{2}]}{\sqrt{\pi} N^{2s} s \Gamma[s]}
\end{eqnarray*}
Bearing in mind that the spectral $K_0$-zeta function is $\zeta_{K_0^{(N)}}(s)=\frac{l}{2\sqrt{\pi}}(N^2)^{\frac{1}{2}-s} \frac{\Gamma[s-\frac{1}{2}]}{\Gamma[s]}$, the regularized one-loop shift due to mass renormalization becomes:
\[
\Delta E_2(\phi_K^{(N)})[s]  = - \frac{N(N+1)}{2\sqrt{\pi}} (N^2)^{-\frac{1}{2}-s} \frac{\Gamma[s+\frac{1}{2}]}{\Gamma[s]}
\]
The Laurent expansions around the pole $s=-\frac{1}{2}$ of both $\Delta E_1(\phi_K^{(N)})[s]$ and $\Delta E_2(\phi_K^{(N)})[s]$,
\begin{eqnarray*}
\textstyle \Delta E_1(\phi_K^{(N)})[s] &=& \sum_{r=1}^{N-1} \left[- \frac{r}{2\pi} \frac{1}{s+\frac{1}{2}} + \frac{r}{2\pi} \left( \gamma +\log (N^2-r^2)+ \psi(-{\textstyle\frac{1}{2}}) - {}_2F_1'[{\textstyle\frac{1}{2},0,\frac{3}{2},-\frac{r^2}{N^2-r^2}}] \right) \right] - \\ && - \frac{N}{2\pi} \frac{1}{s+{\textstyle\frac{1}{2}}} + \frac{N}{2\pi} \left( -2 +\gamma + 2 \log N + \psi(-{\textstyle\frac{1}{2}}) \right) + {\cal O}(s+{\textstyle\frac{1}{2}}) = \\ &=& -\frac{N(N+1)}{4 \pi} \frac{1}{s+\frac{1}{2}} + \left[ \frac{N(N+1)}{4\pi} \left( \gamma+\psi(-{\textstyle \frac{1}{2}}) \right) -\frac{N}{\pi} + \frac{1}{2\pi} N \log N^2 + \right.\\ && \left.+ \frac{1}{2\pi} \sum_{r=1}^{N-1} r \log(N^2-r^2) - \frac{1}{2\pi}\sum_{r=1}^{N-1} r \cdot {}_2F_1'[{\textstyle \frac{1}{2},0,\frac{3}{2},-\frac{r^2}{N^2-r^2}}] \right] + {\cal O}(s+\frac{1}{2})\\
\textstyle \Delta E_2(\phi_K^{(N)})[s]&=&  \frac{N(N+1)}{4\pi} \frac{1}{s+\frac{1}{2}} \textstyle - \frac{N(N+1)}{4\pi} \left(\gamma+ \log N^2 + \psi(-\frac{1}{2}) \right) + {\cal O}(s+\frac{1}{2})\\
\end{eqnarray*}
pave the way to follow a cautious path falling in the physical limit:
\begin{eqnarray}
\Delta E(\phi_K^{(N)}) &= & \frac{1}{\hbar m_d} \lim_{s\rightarrow -\frac{1}{2}} \left[ \Delta E_1(\phi_K^{(N)})[s] + \Delta E_2(\phi_K^{(N)})[s]\right] = -\frac{N}{\pi} + \frac{1}{2\pi} N \log N^2 - \frac{N(N+1)}{4\pi}\log N^2 +\nonumber \\ && + \frac{1}{2\pi} \sum_{r=1}^{N-1} r \log(N^2-r^2)-\frac{1}{2\pi} \sum_{r=1}^{N-1} r \cdot {}_2F_1'[{\textstyle \frac{1}{2},0,\frac{3}{2},-\frac{r^2}{N^2-r^2}}]\label{olmszf} \quad .
\end{eqnarray}
 Here, ${}_2^{} F_1'[a,b,c,z]$ stands for the derivative of the hypergeometric function with respect to the second argument; $\gamma$ is the Euler number, and $\psi(z)$ denotes the digamma function. Note that $\Delta E_1(\phi_K^{(N)})[s]$ and $\Delta E_2(\phi_K^{(N)})[s]$ have a simple pole at $s=-\frac{1}{2}$ with opposite strengths. Therefore, the limit in (\ref{olmszf}) is well defined and provides us with a finite answer for the quantum correction to the kink mass. Despite appearances, formula (\ref{olmszf}) gives the same results as the CCG and DHN formulas, as one can check from standard Tables of special functions.

\subsubsection{The $K$-heat kernel asymptotics}

The main lesson of our previous calculations is that the $\sigma=N$-kinks are very special because the kink wells
are transparent. This fortunate fact allows the computation of the kink semiclassical mass by means of relatively simple formulas extracted from the fairly complete spectral information about the corresponding Schr\"odinger
operators. Dealing with more exotic kinks and self-dual vortices of various types, see \cite{Alonso2006} and \cite{Mateos2009} for reviews, a method has been developed that only requires the knowledge of the potential well $V(x)$ in the computation of one-loop mass shifts of low dimensional topological defects. Starting from the Gilkey-DeWitt-Avramidi high-temperature expansion of the heat function kernel \cite{Gilkey1984,Roe1988,Vassilevich2003,DeWitt1965,Avramidi}, the idea is to express, via the Mellin transform, the energy shifts $\Delta E(\phi_K)$ as a truncated series whose coefficients are the Seeley coefficients of the $K$-heat function. We stress again that this strategy only requires knowledge of the kink potential well $V(x)$, which encodes all the spectral information; the Seeley coefficients are integrals over the real line of the Seeley densities, built from polynomials in $V(x)$ and its derivatives. Therefore, the virtue of this procedure is the universal application to any (1+1)-dimensional one-component scalar field theory model supporting kinks, see \cite{Alonso2002} and \cite{Alonso2006}. The asymptotic approach, however, requires arduous calculations which can be alleviated by using a symbolic software platform, although it has been successfully generalized to (1+1)-dimensional $N$-component scalar field theory \cite{Alonso2004,Alonso2002b,Mateos2009} and even to (1+2)-dimensional Abelian Higgs models \cite{Alonso2005,Alonso2008}. Briefly, the key formula giving the one-loop kink mass shift has the form of a truncated series{\footnote{A similar formula also works for planar self-dual vortices.}}:
\begin{equation}
\Delta E(\phi_K,N_t)=-\frac{1}{8\pi} \sum_{n=2}^{N_t} {c}_n(K)(v^2)^{1-n}\gamma[n-1,\beta_0 v^2]-\frac{1}{2\sqrt{\pi\beta_0}} \quad ,
\label{numericalcorrection}
\end{equation}
where $\gamma[a,b]$ is the incomplete gamma function, $c_n(K)$ are the associated Seeley coefficients of the $K$-heat function expansion, $N_t\in \mathbb{N}$ is the truncation order of the series, and $\beta_0$ is a parameter that freely sets the integration range in the Mellin transform in such a way that the outcome will be optimized. We test
 the method by applying this formula to the $\sigma=N$-kinks, a case where we know the exact result. In fact, because we know the $K^{(N)}$-heat and $K_0^{(N)}$ exactly  we merely expand (\ref{Heat0}) in powers of $\beta$
\begin{equation}
h_{K^{(N)\perp}}(\beta) - h_{K_0}(\beta) =\frac{e^{-N^2 \beta}}{2\sqrt{\pi}} \sum_{n=1}^\infty \frac{2^{n+1}}{(2n-1)!!} \left( \sum_{r=1}^N r^{2n-1}\right) \beta^{n-\frac{1}{2}}-1=\frac{e^{-\beta {v}^2}}{2\sqrt{\pi}} \sum_{n=1}^\infty {c}_n(K^{(N)})\beta^{n-\frac{1}{2}}-1 \label{heatseries} \quad ,
\end{equation}
using the very well known series expansion of the Error function, to immediately identify the Seeley coefficients{\footnote{Note that in formula (\ref{numericalcorrection}) we do not need the first two coefficient: $c_0(K)$ and $c_1(K)$. We have shown in previous works that the contribution of $c_0$ is removed
by zero-point renormalization whereas the $c_1$ term disappears after mass renormalization.}}:
\begin{equation}
{c}_n(K^{(N)}) =\frac{2^{n+1}}{(2n-1)!!} \left( \sum_{r=1}^N r^{2n-1} \right) \quad .
\label{genericSeeleyCoef}
\end{equation}
Plugging these coefficients (\ref{genericSeeleyCoef}) into expression (\ref{numericalcorrection}) and taking large enough $N_t$ and $\beta_0$, we reproduce the results for the one-loop kink mass shifts given in Table 2.

In general there is no exact information on the $K$-heat function. Nevertheless, the Seeley coefficients $c_n(K)$ can
be identified from the $K$-heat kernel expansion:
\begin{equation}
{c}_n(K)=\int_\mathbb{R}  dx \, c_n(x,x)=\int_\mathbb{R}  dx \, {^{(0)}C}_n(x) \qquad .
\label{seeleycoef}
\end{equation}
The Seeley densities $c_n(x,x)$ can be obtained by solving the recurrence relations -coming from the series expansion of the fundamental solution of the $K$-heat equation-
\begin{equation}
{^{(k)} C}_n(x) =\frac{1}{n+k} \left[ \rule{0cm}{0.6cm} \right.
{^{(k+2)} C}_{n-1}(x) - \sum_{j=0}^k {k \choose j}
\frac{\partial^j V}{\partial x^j}\, \, {^{(k-j)}
C}_{n-1}(x) \left. \rule{0cm}{0.6cm} \right]
\label{capitalAcoefficients}
\end{equation}
with \lq\lq initial" (infinite temperature) conditions: ${^{(k)} C}_0(x)= \delta^{k0}$.

In order to test the asymptotic approach we now compute (\ref{numericalcorrection}) from the coefficients obtained as solutions of the recurrence relations (\ref{seeleycoef}) and (\ref{capitalAcoefficients}). We shall choose a truncation order $N_t=40$, e.g., in the fourth member $N=4$ of the hierarchy: $U^{(4)}(\phi)=\frac{9}{8} [1-2 \cos(\frac{2}{3} \arcsin \phi)]^4, |\phi|\leq 1$. The $K^{(4)}$ Seeley coefficients arising from the recurrence relations are shown in Table 3.

\begin{table}[ht]
\hspace{1cm}\begin{tabular}{|c|r|} \hline
$n$ & $c_n(K^{(4)})$ \hspace{0.3cm} \\ \hline
$1$ & $40.0000$ \\
$2$ & $266.6667$ \\
$3$ & $1386.6667$ \\
$4$ & $5699.0476$ \\
$5$ & $19121.4392$ \\
$6$ & $53853.5835$ \\
$7$ & $130166.9684$ \\
$8$ & $274845.7626$ \\
$9$ & $514360.2887$ \\
$10$ & $863458.4292$ \\
\hline
\end{tabular} \hspace{0.5cm}
\begin{tabular}{|c|r|} \hline
$n$ & $c_n(K^{(4)})$ \hspace{0.3cm} \\ \hline
$11$ & $1313320.5922$ \\
$12$ & $1825331.1803$ \\
$13$ & $2335057.9936$ \\
$14$ & $2766565.6064$ \\
$15$ & $3052196.9105$ \\
$16$ & $3150326.7363$ \\
$17$ & $3054683.3061$ \\
$18$ & $2792761.2600$ \\
$19$ & $2415316.3095$ \\
$20$ & $1981777.3296$ \\
\hline
\end{tabular} \hspace{0.5cm}
\begin{tabular}{|c|r|} \hline
$n$ & $c_n(K^{(4)})$ \hspace{0.3cm} \\ \hline
$21$ & $1546743.9642$ \\
$22$ & $1151061.4774$ \\
$23$ & $818531.0868$ \\
$24$ & $557297.1794$ \\
$25$ & $363948.9646$ \\
$26$ & $228360.0592$ \\
$27$ & $137877.7460$ \\
$28$ & $80219.7711$ \\
$29$ & $45035.6583$ \\
$30$ & $24426.1190$ \\
\hline
\end{tabular} \hspace{0.5cm}
\begin{tabular}{|c|r|} \hline
$n$ & $c_n(K^{(4)})$ \hspace{0.3cm} \\ \hline
$31$ & $12813.7015$ \\
$32$ & $6508.5467$ \\
$33$ & $3204.2076$ \\
$34$ & $1530.3678$ \\
$35$ & $709.7358$ \\
$36$ & $319.8809$ \\
$37$ & $140.2218$ \\
$38$ & $59.8280$ \\
$39$ & $24.8636$ \\
$40$ & $10.0713$ \\
\hline
\end{tabular}
\caption{\small Seeley coefficients of the $K^{(4)}$-heat function expansion.}
\end{table}

These data coincide exactly with those obtained from the analytical formula (\ref{genericSeeleyCoef}) for $N=4$:
\[
{c}_n(K^{(4)}) =\frac{2^{n+1}}{(2n-1)!!} \left( 1^{2n-1} + 2^{2n-1} + 3^{2n-1} + 4^{2n-1} \right)\quad .
\]
The $40$ Seeley coefficients, obtained either way, are now plugged into formula (\ref{numericalcorrection}). The tuning of the $\beta_0$ parameter to the value $\beta_0=1.3125$ offers an optimum estimation of the one-loop kink mass shift
\[
\Delta E(\phi_K^{(4)}) \approx -1.58003
\]
to be compared, e.g., with the answer given by the CCG formula. The link between $N_t$ and $\beta_0$ can be understood by looking at Figures 3(b) and 3(c): there, several partial sums, as well as the total series, are plotted  as functions of $\beta$. For each $N_t$-order there is a point $\beta_0$ in the $\beta$-half-line where the partial sum is closer to the exact value. The further we can push $\beta_0$ towards infinity, the larger part of the function we can pick in the integral, and the optimum value is reached.

\subsection{The $\sigma \notin \mathbb{N}$ case}

In this section we shall apply the generalized DHN formula and the asymptotic approach to compute the one-loop (\ref{kinkN2})-kink mass shifts when $\sigma\notin \mathbb{N}$. In this case $r^{(\sigma)}(k)$, given in (\ref{asdata}), are not null and, moreover, there is no half-bound state in the spectrum of $K^{(\sigma)}$.

\subsubsection{The generalized DHN formula}

In \cite{Alonso2004} the extension of formula (\ref{dhn}) to this more general situation has been described thoroughly. The \lq\lq generalized" DHN formula reads:
\begin{equation}
\Delta E(\phi_K)= \frac{1}{2}\sum_{j=0}^{l-1} \omega_j + \frac{1}{2} s_l \omega_l- \frac{v}{4}  + \frac{1}{2\pi} \int_0^\infty dk \frac{d \delta}{d k} \sqrt{k^2 + v^2} + \frac{1}{4\pi} \left<V(x) \right> -\frac{1}{8\pi}  \left<V(x) \right>  \int_{-\infty}^\infty \frac{dk}{\sqrt{k^2+v^2}}
\label{gdhn}
\end{equation}
where $\omega_j^2$ are the bound-state eigenvalues of $K=-\frac{d^2}{dx^2}+v^2+V(x)$ and $s_l$ is equal to $\frac{1}{2}$ if the higher eigenvalue $\omega_l^2=v^2$  lies in the continuous spectrum threshold,  but it is otherwise one: $\omega_l^2<v^2$. The new ingredient in the formula encodes the imbalance between the half-bound states of $K$ and $K_0$.

The application of formula (\ref{gdhn}) to the $\sigma\notin \mathbb{N}$ models only requires knowledge of the spectral density derived from (\ref{asdata}) and the consequent total phase shift:
\begin{equation}
\frac{d \delta^{(\sigma)}}{d k} = \psi(-ik)-\psi(-\sigma-ik)-\psi(1+\sigma-ik) + \frac{\psi(ik)\Gamma^2[ik] \Gamma^2[1-ik]-\psi(1-ik)\Gamma^2[-\sigma] \Gamma^2[1+\sigma]}{\Gamma^2[1-ik]\Gamma^2[ik]-\Gamma^2[-\sigma]\Gamma^2[1+\sigma]}\label{specden} \quad .
\end{equation}
Plugging (\ref{specden}) into (\ref{gdhn}) for the $\sigma\notin \mathbb{N}$-kinks we obtain:
\[
\Delta E(\phi_K^{(\sigma)}) =\frac{1}{2} \sum_{j=0}^{I[\sigma]} \sqrt{j(2\sigma-j)} - \frac{\sigma(\sigma+1)}{2\pi} - \frac{\sigma}{4} + \frac{1}{2\pi} \int_0^\infty dk \left[ \frac{d \delta^{(\sigma)}}{d k} \sqrt{k^2+\sigma^2} + \frac{\sigma(1+\sigma)}{\sqrt{k^2+\sigma^2}} \right]
\]
 Owing to the complexity of the derivative of the phase shifts, exact integration of this formula is not possible. Thus, we extract the information by means of numerical integration, offering the following results:
\begin{table}[ht]
\centerline{\begin{tabular}{|c|c|} \\[-0.6cm] \hline
$\sigma$ & $\Delta E[\phi_{\rm K}^{(\sigma)}]$ \\ \hline
${\bf 0.0}$ & $ 0.00000 $ \\
$0.01$ & $ -0.0040164 $ \\
$0.1$ & $ -0.037024 $ \\
$0.2$ & $ -0.070414 $ \\
$0.3$ & $ -0.102158 $ \\
$0.4$ & $ -0.133080 $ \\
$0.5$ & $ -0.163637 $ \\
$0.6$ & $ -0.194114 $ \\
$0.7$ & $ -0.224700 $ \\
$0.8$ & $ -0.255533 $ \\
$0.9$ & $ -0.286711 $ \\
$0.99$ & $ -0.315129 $ \\
\hline
\end{tabular} \hspace{0.1cm}
\begin{tabular}{|c|c|} \\[-0.6cm] \hline
$\sigma$ & $\Delta E[\phi_{\rm K}^{(\sigma)}]$ \\ \hline
${\bf 1.0}$ & $ -0.31831 $ \\
$1.01$ & $ -0.321495 $ \\
$1.1$ & $ -0.350388 $ \\
$1.2$ & $ -0.382994$ \\
$1.3$ & $ -0.416163 $ \\
$1.4$ & $ -0.449927 $ \\
$1.5$ & $ -0.484311 $ \\
$1.6$ & $ -0.519340 $ \\
$1.7$ & $ -0.555028 $ \\
$1.8$ & $ -0.591393 $ \\
$1.9$ & $ -0.628449 $ \\
$1.99$ & $ -0.662399 $ \\
\hline
\end{tabular} \hspace{0.1cm}
\begin{tabular}{|c|c|} \\[-0.6cm] \hline
$\sigma$ & $\Delta E[\phi_{\rm K}^{(\sigma)}]$ \\ \hline
${\bf 2.0}$ & $ -0.666254 $  \\
$2.01$ & $ -0.670018 $ \\
$2.1$ & $ -0.704683 $ \\
$2.2$ & $ -0.743871 $ \\
$2.3$ & $ -0.783797 $ \\
$2.4$ & $ -0.824449 $ \\
$2.5$ & $ -0.865861$ \\
$2.6$ & $ -0.908014 $ \\
$2.7$ & $ -0.950919 $ \\
$2.8$ & $ -0.986907 $ \\
$2.9$ & $ -1.03902 $ \\
$2.99$ & $-1.07967 $ \\
\hline
\end{tabular} \hspace{0.1cm}
\begin{tabular}{|c|c|} \\[-0.6cm] \hline
$\sigma$ & $\Delta E[\phi_{\rm K}^{(\sigma)}]$ \\ \hline
${\bf 3.0}$ & $ -1.08451 $  \\
$3.01$ & $ -1.08878 $  \\
$3.1$ & $ -1.13019 $  \\
$3.2$ & $ -1.17694 $  \\
$3.3$ & $ -1.22446 $  \\
$3.4$ & $ -1.27276 $  \\
$3.5$ & $ -1.32184 $  \\
$3.6$ & $ -1.37171 $  \\
$3.7$ & $ -1.42237 $  \\
$3.8$ & $ -1.47381 $  \\
$3.9$ & $ -1.52605 $  \\
$3.99$ & $ -1.57374 $  \\ \hline
\end{tabular}}
\caption{One-loop parent kink mass shifts for several values of $\sigma \in [0.0,4.0]$ estimated by applying the generalized DHN formula.}
\end{table}
\begin{figure}[ht]
\centerline{
\includegraphics[height=4.cm]{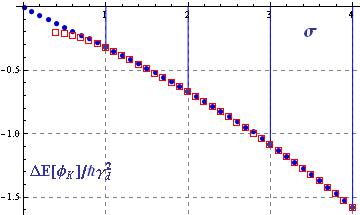} \hspace{1cm}
\includegraphics[height=4.cm]{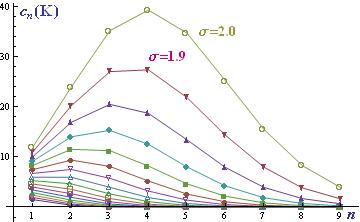} }
\caption{\small Graphical representation of: (a) one-loop kink mass shifts in the range: $\sigma\in [0.0,4.0]$. (b) Seeley coefficients within the same range.}
\end{figure}

The values depicted in Table 4 have been drawn in Figure 4 as solid points. The figure shows a remarkable pattern: the continuity of $\Delta E[\phi_{\rm K}^{(\sigma)}]$  as a function of the parameter $\sigma$. The generalized DHN formula (derived in the mode-number cut-off regularization procedure plus a $\frac{1}{2}$-weight on the half-bound states coming from the 1D Levinson theorem) guarantees this behaviour. Had we applied the original DHN formula for all values of $\sigma\in \mathbb{R}$ we would have found discontinuities at integers values of $\sigma$. Note also that $\Delta E[\phi_{\rm K}^{(\sigma)}]/m_d\hbar$ tends to zero when $\sigma$ vanishes, as should be the case, see Figure 4(a).

\subsubsection{The asymptotic approach}

Finally, we shall apply the asymptotic approach in the $\sigma\notin\mathbb{N}$ models. In this case, computation of the Seeley coefficients $c_n(K^{(\sigma)})$ in (\ref{numericalcorrection}) can only be achieved by solving the recurrence relations (\ref{capitalAcoefficients}). In Reference \cite{Alonso2011} a symbolic program code has been developed that performs this laborious task. We use this code to calculate the Seeley coefficients, adapted to the present models. A graphical representation of these coefficients is shown in Figure 4(b) for different values of the $\sigma$ parameter  with a separation of 0.1 in the range $[0.4,2.0]$. In general, the larger the value of $\sigma$, the larger the coefficients $c_n(K^{(\sigma)})$. Plugging these coefficients into formula (\ref{numericalcorrection}) we estimate the quantum correction. Frequently, the response of (\ref{numericalcorrection}) has been evaluated by employing a series involving ten terms, i.e., $N_t=10$, although this can vary along the values of the parameter $\sigma$. The results obtained are shown in Table 5.

\begin{table}[ht]
\centerline{\begin{tabular}{|c|c|} \\[-0.6cm] \hline
$\sigma$ & $\Delta E[\phi_{\rm K}^{(\sigma)}]$ \\ \hline
${\bf 0.0}$ & $ -- $ \\
$0.01$ & $ -- $ \\
$0.1$ & $ -- $ \\
$0.2$ & $ -- $ \\
$0.3$ & $ -- $ \\
$0.4$ & $ -0.201364 $ \\
$0.5$ & $ -0.211213 $ \\
$0.6$ & $ -0.224572 $ \\
$0.7$ & $ -0.241903 $ \\
$0.8$ & $ -0.263464 $ \\
$0.9$ & $ -0.289269 $ \\
$0.99$ & $ -0.316476 $ \\
\hline
\end{tabular} \hspace{0.1cm}
\begin{tabular}{|c|c|} \\[-0.6cm] \hline
$\sigma$ & $\Delta E[\phi_{\rm K}^{(\sigma)}]$ \\ \hline
${\bf 1.0}$ & $ -0.318321 $ \\
$1.01$ & $ -0.322882 $ \\
$1.1$ & $ -0.352179 $ \\
$1.2$ & $ -0.385393$ \\
$1.3$ & $ -0.416517 $ \\
$1.4$ & $ -0.450447 $ \\
$1.5$ & $ -0.484716 $ \\
$1.6$ & $ -0.519600 $ \\
$1.7$ & $ -0.555152 $ \\
$1.8$ & $ -0.591376 $ \\
$1.9$ & $ -0.628289 $ \\
$1.99$ & $ -0.662103 $ \\
\hline
\end{tabular} \hspace{0.1cm}
\begin{tabular}{|c|c|} \\[-0.6cm] \hline
$\sigma$ & $\Delta E[\phi_{\rm K}^{(\sigma)}]$ \\ \hline
${\bf 2.0}$ & $ -0.666241 $  \\
$2.01$ & $ -0.669769 $ \\
$2.1$ & $ -0.704338 $ \\
$2.2$ & $ -0.743427 $ \\
$2.3$ & $ -0.783236 $ \\
$2.4$ & $ -0.823772 $ \\
$2.5$ & $ -0.865044$ \\
$2.6$ & $ -0.907056 $ \\
$2.7$ & $ -0.949812 $ \\
$2.8$ & $ -0.99332 $ \\
$2.9$ & $ -1.03758 $ \\
$2.99$ & $ -1.07807 $ \\ \hline
\end{tabular} \hspace{0.1cm}
\begin{tabular}{|c|c|} \\[-0.6cm] \hline
$\sigma$ & $\Delta E[\phi_{\rm K}^{(\sigma)}]$ \\ \hline
${\bf 3.0}$ & $ -1.08441 $  \\
$3.01$ & $ -1.08714 $  \\
$3.1$ & $ -1.12839 $  \\
$3.2$ & $ -1.17493 $  \\
$3.3$ & $ -1.22225 $  \\
$3.4$ & $ -1.27033 $  \\
$3.5$ & $ -1.31919 $  \\
$3.6$ & $ -1.36882 $  \\
$3.7$ & $ -1.41924 $  \\
$3.8$ & $ -1.47043 $  \\
$3.9$ & $ -1.52240 $  \\
$3.99$ & $ -1.56984 $  \\
\hline
\end{tabular}}
\caption{One-loop (\ref{kinkN2}) kink mass shifts for several values of $\sigma \in [0.0,4.0]$  estimated by applying the asymptotic approach.}
\end{table}

The values seen in the Table 5 are drawn in Figure 4(a) as empty squares in order to appreciate the concordance between the data obtained upon using the generalized DHN formula and the asymptotic approach. Only a small difference between the data arises for $\sigma\leq 0.6$ and lower. For $\sigma$ small the series in (\ref{numericalcorrection}) converges slowly because of the factor $(\sigma^2)^{1-n}$ in each summand; a larger number of terms is needed to obtain a precise answer. The limitations in the computations of the Seeley coefficients make the asymptotic approach less exact in the range of values of the parameter $\sigma<0.4$.

\section{Conclusions and further comments}

We have constructed a one-parametric family of (1+1)-dimensional one-component scalar field theory models governed by the action (\ref{action}) with the potential (\ref{potentialinPhi}). This family includes two famous members: the sine-Gordon and $\phi^4$ models. The models in this family are characterized by very specific kink wells: the second-order kink fluctuation operators are Schr\"odinger operators with P\"oschl-Teller potential wells. In general, the family of potentials (\ref{potentialinPhi}) is only of ${\cal C}^2(\mathbb{R})$ class as a function of the field $\phi$: Derivatives of order higher than two are discontinuous or divergent at the vacuum points. There exists a similar family of models, with class ${\cal C}^\infty$ potentials, for two or more scalar fields. For example, the (1+1)-dimensional two-component scalar field theory model with a ${\cal C}^\infty(\mathbb{R}^2)$ potential
\[
U(\phi_1,\phi_2)=\frac{1}{8}(4\phi_1^2 +2 \sigma \phi_2^2-1)^2 + 2\sigma^2 \phi_1^2 \phi_2^2
\]
admits the kink solution
\[
\phi_1^{\rm K}(x)=\frac{1}{2} \tanh \overline{x} \hspace{1cm} , \hspace{1cm} \phi_2^{\rm K}(x) =0
\]
whose second-order small fluctuation operator is:
\[
K^{(\sigma)} = \left( \begin{array}{cc} -\frac{d^2}{dx^2}+4-6\,{\rm sech}^2\,\overline{x} & 0 \\ 0 & -\frac{d^2}{dx^2}+\sigma^2-\sigma(\sigma+1) \,{\rm sech}^2\,\overline{x} \end{array} \right)\qquad .
\]
Here, the generic P\"oschl-Teller potential well arises in the study of the orthogonal fluctuations to the kink solution, see \cite{Alonso2004}.

Finally, the spectral problems associated with kink stability analysis in the models studied in this paper are solvable and therefore the one-loop quantum correction to the kink mass in these cases can be computed exactly. By the same token as before, $\Delta E^\bot[\phi_{\rm K}^{(\sigma)}]$, the one-loop kink mass shift due to orthogonal  fluctuations to the kink, is a continuous function of $\sigma$. Note, however, that our methods have been improved
such that the agreement between the results via the generalized DHN formula and the asymptotic method is now much better than eight years ago, see \cite{Alonso2004}.

\section*{ACKNOWLEDGEMENTS}

We gratefully acknowledge partial funding from
the Spanish Ministerio de Educacion y Ciencia (DGICYT) under grant: FIS2009-10546.

\end{document}